\documentclass[usegraphicx,usenatbib]{mn2e}

\usepackage{graphicx}
\usepackage{natbib}
\usepackage{fixltx2e}
\usepackage{mathptmx}
\usepackage{amsmath}
\usepackage{wasysym}
\usepackage{url}
\usepackage{times}
\usepackage{epsfig}
\usepackage{ulem}

\newcommand{\Mpc}{\rm\; Mpc}
\newcommand{\kpc}{\rm\; kpc}

\newcommand{\km}{\rm\; km}

\newcommand{\cm}{\rm\; cm}
\newcommand{\mum}{\hbox{$\rm\; \mu m\,$}}

\newcommand{\yr}{\rm\; yr}
\newcommand{\Gyr}{\rm\; Gyr}

\newcommand{\s}{\rm\; s}

\newcommand{\ks}{\rm\; ks}

\newcommand{\keVpcmcu}{\hbox{$\keV\cm^{-3}\,$}}

\newcommand{\Msun}{\hbox{$\rm\thinspace M_{\odot}$}}

\newcommand{\Msunpyr}{\hbox{$\Msun\yr^{-1}\,$}}

\newcommand{\keV}{\rm\; keV}

\newcommand{\erg}{\rm\; erg}

\newcommand{\ergpcmsqps}{\hbox{$\erg\cm^{-2}\s^{-1}\,$}}

\newcommand{\ergps}{\hbox{$\erg\s^{-1}\,$}}

\newcommand{\kmps}{\hbox{$\km\s^{-1}\,$}}

\newcommand{\kmpspMpc}{\hbox{$\kmps\Mpc^{-1}\,$}}

\newcommand{\Zsun}{\hbox{$\thinspace \mathrm{Z}_{\odot}$}}

\newcommand{\muG}{\hbox{$\rm\thinspace {\mu}\mathrm{G}$}}
\newcommand{\amin}{\rm\; arcmin}

\newcommand{\asec}{\rm\; arcsec}

\newcommand{\emm}{\hbox{$\cm^{-5}\,$}}
\newcommand{\sqremm}{\hbox{$\cm^{-5/2}\,$}}
\newcommand{\empasecsq}{\hbox{$\emm\asec^{-2}\,$}}

\newcommand{\pseudoP}{\hbox{$\keV\sqremm\asec^{-2}\,$}}

\newcommand{\psqcm}{\hbox{$\cm^{-2}\,$}}

\newcommand{\pcmcu}{\hbox{$\cm^{-3}\,$}}

\begin{document}

\title[The merging galaxy group RXJ0751.3+5012]{The bow shock, cold fronts and disintegrating cool core in the merging galaxy group RXJ0751.3+5012}
\author[]  
    {\parbox[]{7.in}{H.~R. Russell$^{1,2,3}$\thanks{E-mail: 
          hrr27@ast.cam.ac.uk}, A.~C. Fabian$^3$, B.~R. McNamara$^{1,4,5}$, A.~C. Edge$^{2}$, J.~S. Sanders$^{6}$, P.~E.~J. Nulsen$^5$, S.~A. Baum$^{7}$, M. Donahue$^8$ and C.~P. O'Dea$^{9}$\\ 
    \footnotesize 
    $^1$ Department of Physics and Astronomy, University of Waterloo, Waterloo, ON N2L 3G1, Canada\\
    $^2$ Department of Physics, Durham University, Durham DH1 3LE\\
    $^3$ Institute of Astronomy, Madingley Road, Cambridge CB3 0HA\\
    $^4$ Perimeter Institute for Theoretical Physics, Waterloo, Canada\\ 
    $^5$ Harvard-Smithsonian Center for Astrophysics, 60 Garden Street, Cambridge, MA 02138, USA\\
    $^6$ Max-Planck-Institute f\"ur extraterrestrische Physik, 85748 Garching, Germany\\
    $^7$ Chester F. Carlson Center for Imaging Science, Rochester Institute of Technology, Rochester, NY 14623, USA\\
    $^8$ Department of Physics and Astronomy, Michigan State University, East Lansing, MI 48824, USA\\
    $^9$ Department of Physics, Rochester Institute of Technology, Rochester, NY 14623, USA\\
  }
}

\maketitle

\begin{abstract}
We present a new \textit{Chandra} X-ray observation of the off-axis galaxy group merger RXJ0751.3+5012.  The hot atmospheres of the two colliding groups appear highly distorted by the merger.  The images reveal arc-like cold fronts around each group core, produced by the motion through the ambient medium, and the first detection of a group merger shock front.  We detect a clear density and temperature jump associated with a bow shock of Mach number $M=1.9\pm0.4$ ahead of the northern group.  Using galaxy redshifts and the shock velocity of $1100\pm300\kmps$, we estimate that the merger axis is only $\sim10^{\circ}$ from the plane of the sky.  From the projected group separation of $\sim90\kpc$, this corresponds to a time since closest approach of $\sim0.1\Gyr$.  The northern group hosts a dense, cool core with a ram pressure stripped tail of gas extending $\sim100\kpc$.  The sheared sides of this tail appear distorted and broadened by Kelvin-Helmholtz instabilities.  We use the presence of this substructure to place an upper limit on the magnetic field strength and, for Spitzer-like viscosity, show that the development of these structures is consistent with the critical perturbation length above which instabilities can grow in the intragroup medium.  The northern group core also hosts a galaxy pair, UGC4052, with a surrounding IR and near-UV ring $\sim40\kpc$ in diameter.  The ring may have been produced by tidal stripping of a smaller galaxy by UGC4052 or it may be a collisional ring generated by a close encounter between the two large galaxies.
\end{abstract}



\section{Introduction}
\label{sec:intro}

\begin{figure*}
\begin{minipage}{\textwidth}
\centering
\includegraphics[width=0.48\columnwidth]{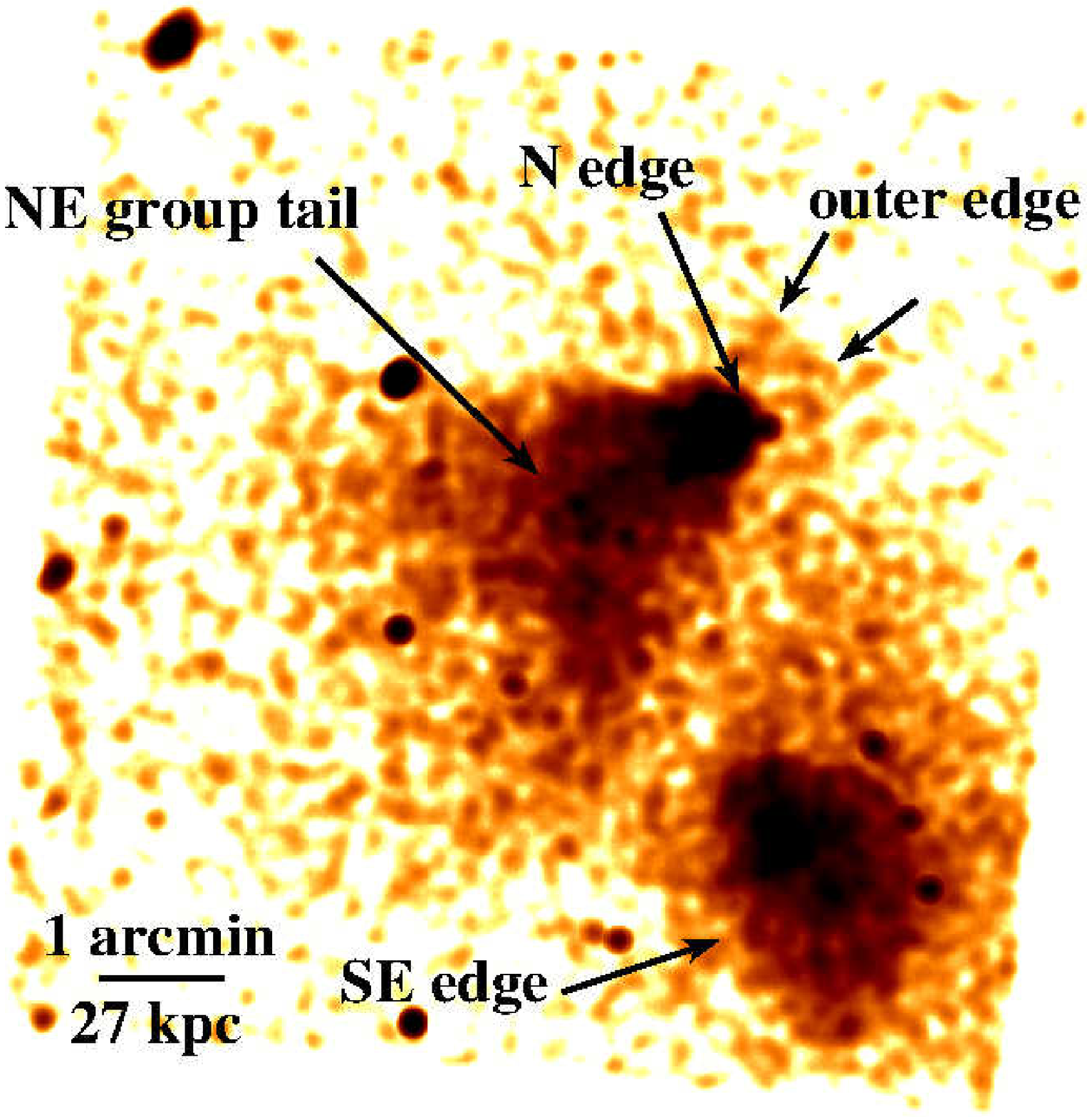}
\includegraphics[width=0.48\columnwidth]{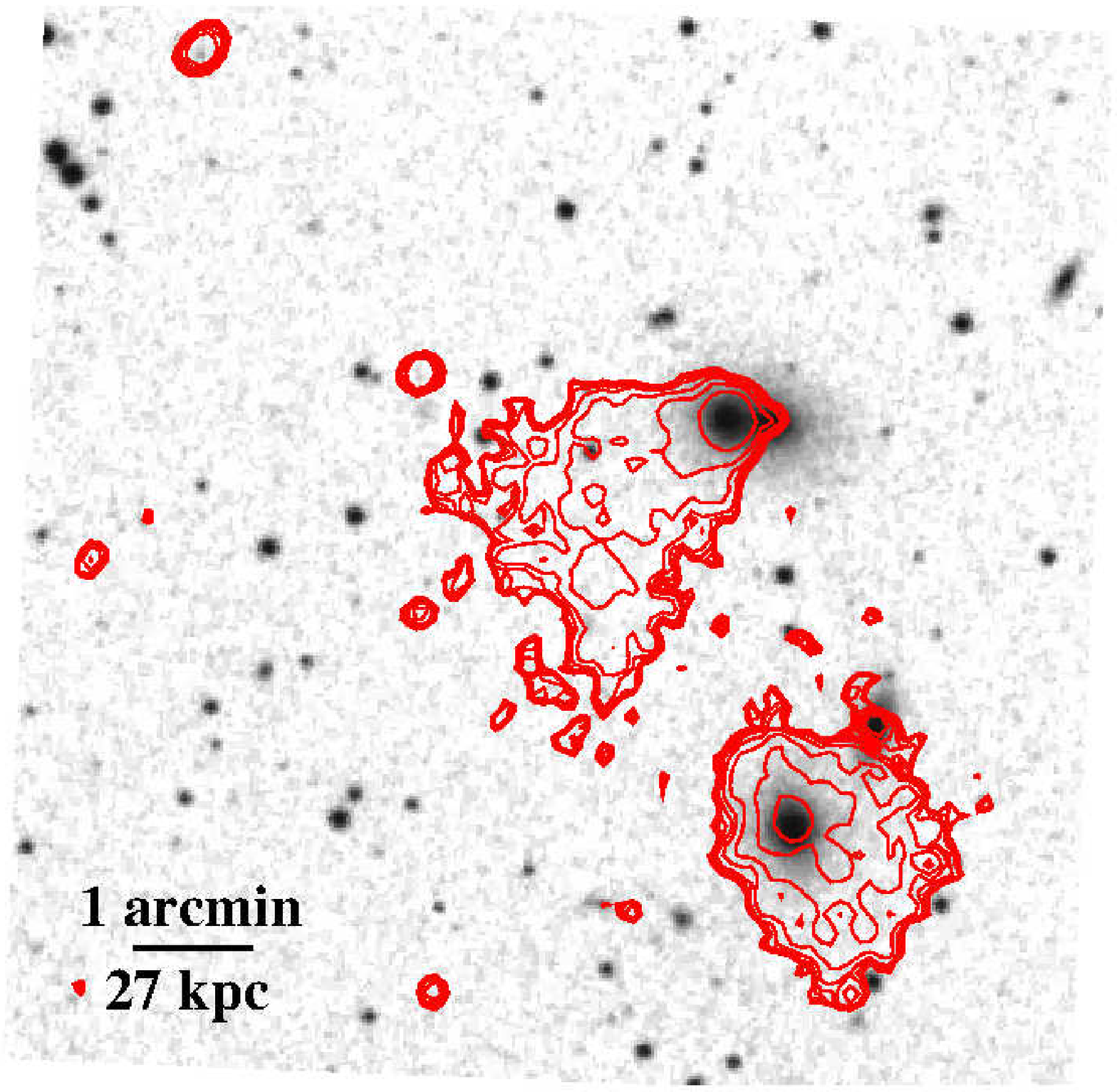}
\caption{Left: Exposure-corrected image in the $0.3-4.0\keV$ energy band smoothed with a 2D Gaussian $\sigma=7\asec$ (North is up and East is to the left).  The arrows indicate a surface brightness edge ahead of the NE group. Right: DSS optical image covering the same field of view with X-ray contours.}
\label{fig:sbimage}
\end{minipage}
\end{figure*}

\begin{figure}
\centering
\includegraphics[width=0.95\columnwidth]{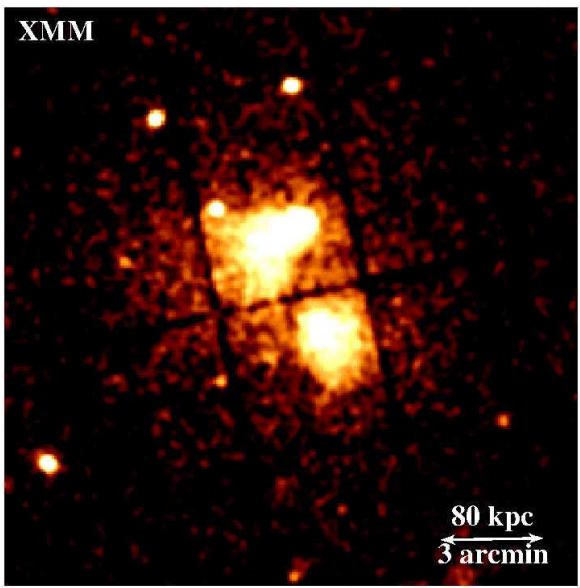}
\caption{XMM-Newton MOS and PN merged image smoothed with a 2D Gaussian $\sigma=12\asec$ to show the large-scale environment of the galaxy groups.}
\label{fig:xmmimage}
\end{figure}

A large fraction of the baryons in the nearby Universe resides in
galaxy groups with X-ray luminosities $\sim10^{41}-10^{43}\ergps$ and
temperatures $\sim0.3-2\keV$ (eg. \citealt{Mulchaey93};
\citealt{Ponman93}; \citealt{Mulchaey00}).  Galaxy groups are much
more representative hosts than rarer rich cluster systems
(eg. \citealt{Geller83}; \citealt{Tully87}) and important for understanding the
gravitational and thermal evolution of most of the matter in the
Universe.  The effects of non-gravitational heating, such as supernova
explosions, AGN feedback and merger shocks, are expected to be more
pronounced in lower mass systems as the energy input from these
sources is comparable to the binding energy of the group (eg.
\citealt{Ponman96}; \citealt{Ponman99}; \citealt{Helsdon00}).  Major
and minor group mergers, and their subsequent relaxation, govern the
formation of the largest-scale structures and can have a substantial
impact on their constituent galaxies. However, studies of galaxy group
mergers have been limited due to their faint X-ray emission and low galaxy densities and richer cluster
mergers have instead been the focus of attention (but see
eg. \citealt{Kraft04,Kraft06}; \citealt{Machacek10,Machacek11}; \citealt{Kraft11}).  

Deep \textit{Chandra} X-ray observations of the hot gas substructures
generated by mergers have proved crucial to our understanding of the
dynamics and thermal evolution.  Narrow surface brightness edges are
observed where the gas density drops across the discontinuity and the
gas temperature increases so that the pressure is continuous
(\citealt{Markevitch00}; \citealt{Vikhlinin01};
\citealt{Markevitch07}).  These `cold fronts' mark the boundary of a
cool, dense core of gas that is moving through the warmer ambient
medium.  The ram pressure experienced by the central gas peak in a
major merger produces a sharp cold front along the leading interface
of the dense core (eg. \citealt{Vikhlinin01}).  Sloshing cold fronts
can occur when a minor merger sets a dense core oscillating in the
gravitational potential well (eg. \citealt{Markevitch01};
\citealt{AscasibarMarkevitch06}).  Shock fronts are also expected in
mergers but there are only a handful of detections in clusters with a
clear density and temperature increase
(eg. \citealt{Markevitch02,Markevitch05}; \citealt{Russell10};
\citealt{Macario11}).  Shock front detections provide measurements of
the gas bulk velocities in the plane of the sky and constrain the
merger geometry.  They are therefore key observational tools for studying
the merger history (eg. \citealt{Markevitch07}).  The detailed
structure of cold fronts has also been used to study the relatively
unknown transport processes in the hot atmospheres of groups.
Observations of Kelvin-Helmholtz instabilities along cold fronts, or
their absence, can provide a measure of the gas viscosity and magnetic
field strength in the ICM (eg. \citealt{VikhlininBfield01};
\citealt{Roediger11}; \citealt{Zuhone11}).  The sharp temperature and
density changes across cold fronts also suggests that thermal
conduction and diffusion are strongly suppressed
(eg. \citealt{Ettori00}; \citealt{Vikhlinin01}).

X-ray observations with \textit{ROSAT} and \textit{XMM-Newton} of the
nearby system RXJ0751.3+5012 ($z=0.022$; \citealt{Ebeling98}; \citealt{Watson09}) revealed two large galaxy
groups with disturbed morphologies, which are in the process of
merging.  The archival \textit{XMM-Newton} observation showed surface
brightness edges around the two groups, each $\sim100\kpc$ across, and
a long tail of stripped material extending behind the northern group
core.  The northern group core hosts luminous optical line emission,
star formation and a striking IR bright ring of emission surrounding
the two central group galaxies (\citealt{Quillen08};
\citealt{ODea08}).  Here we present a new deep \textit{Chandra}
observation of the off-axis group merger RXJ0751.3+5012 studying cold
front and shock front structure, ram pressure stripping and turbulent
mixing in detail in the intragroup medium.  In Section
\ref{sec:chandraanalysis} we present the imaging and spectroscopic
analysis of the Chandra data and identify several surface brightness
edges associated with cold fronts and a bow shock.  The structure of
the bow shock is analysed in Section \ref{sec:shock} and the Mach
number is calculated.  In Sections \ref{sec:core} and \ref{sec:tail}
we discuss the formation of turbulent instabilities around the NE
group core and ram pressure stripped tail and determine an upper limit
on the magnetic field strength.  Finally, in Section
\ref{sec:highzcluster} we show the detection of additional extended
emission in the ACIS-I field of view likely corresponding to the
galaxy cluster WHL\,J075052.4+500252 at $z=0.417$.

We assume $H_0=70\kmpspMpc$, $\Omega_m=0.3$ and $\Omega_\Lambda=0.7$,
translating to a scale of $0.45\kpc$ per arcsec at the redshift
$z=0.022$ of RXJ0751.3+5012.  All errors are $1\sigma$ unless
otherwise noted.

\section{\textit{Chandra} data analysis}
\label{sec:chandraanalysis}

\subsection{Data reduction}
\label{sec:datareduction}

\begin{figure*}
\begin{minipage}{\textwidth}
\centering
\includegraphics[width=0.32\columnwidth]{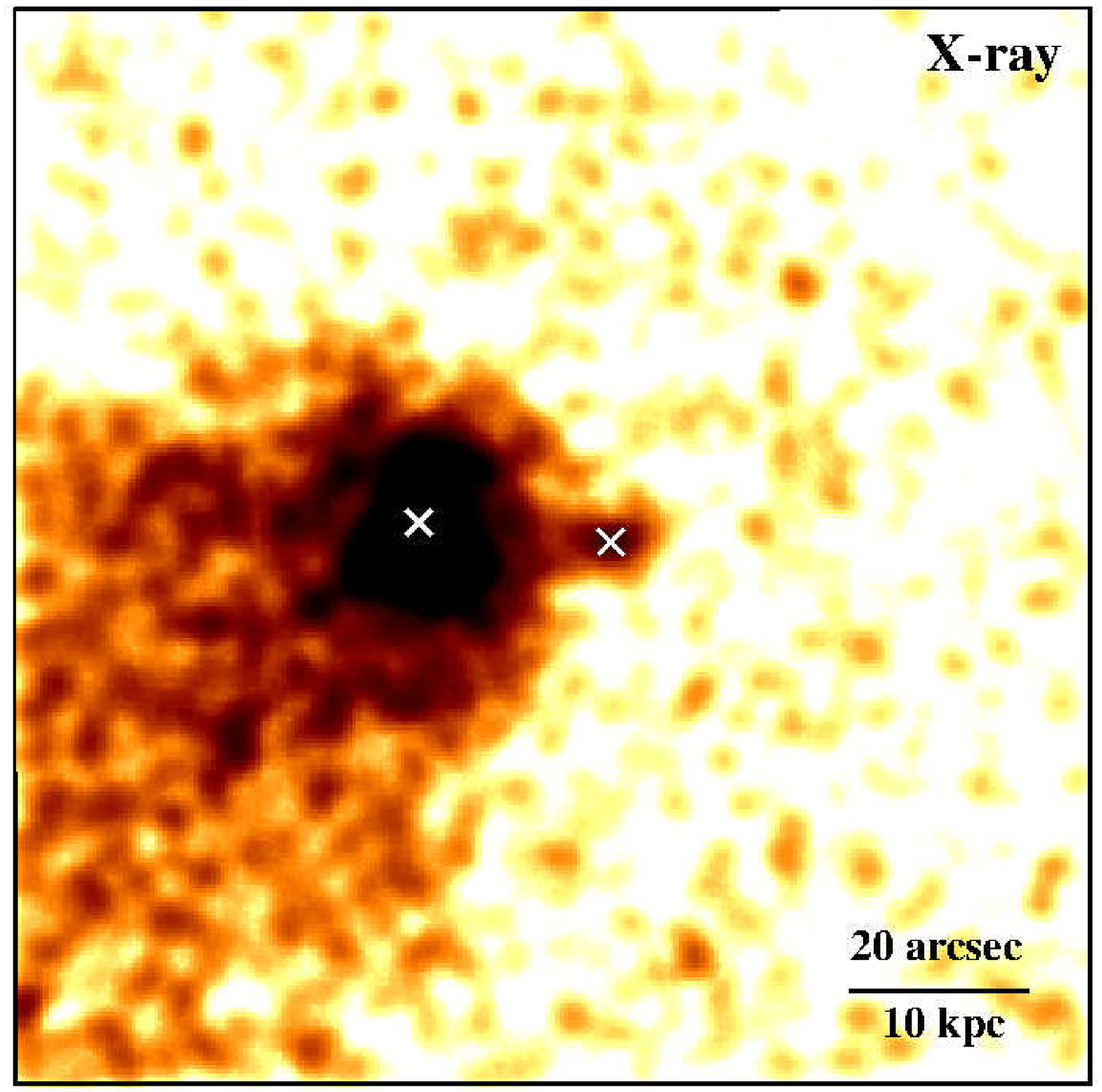}
\includegraphics[width=0.32\columnwidth]{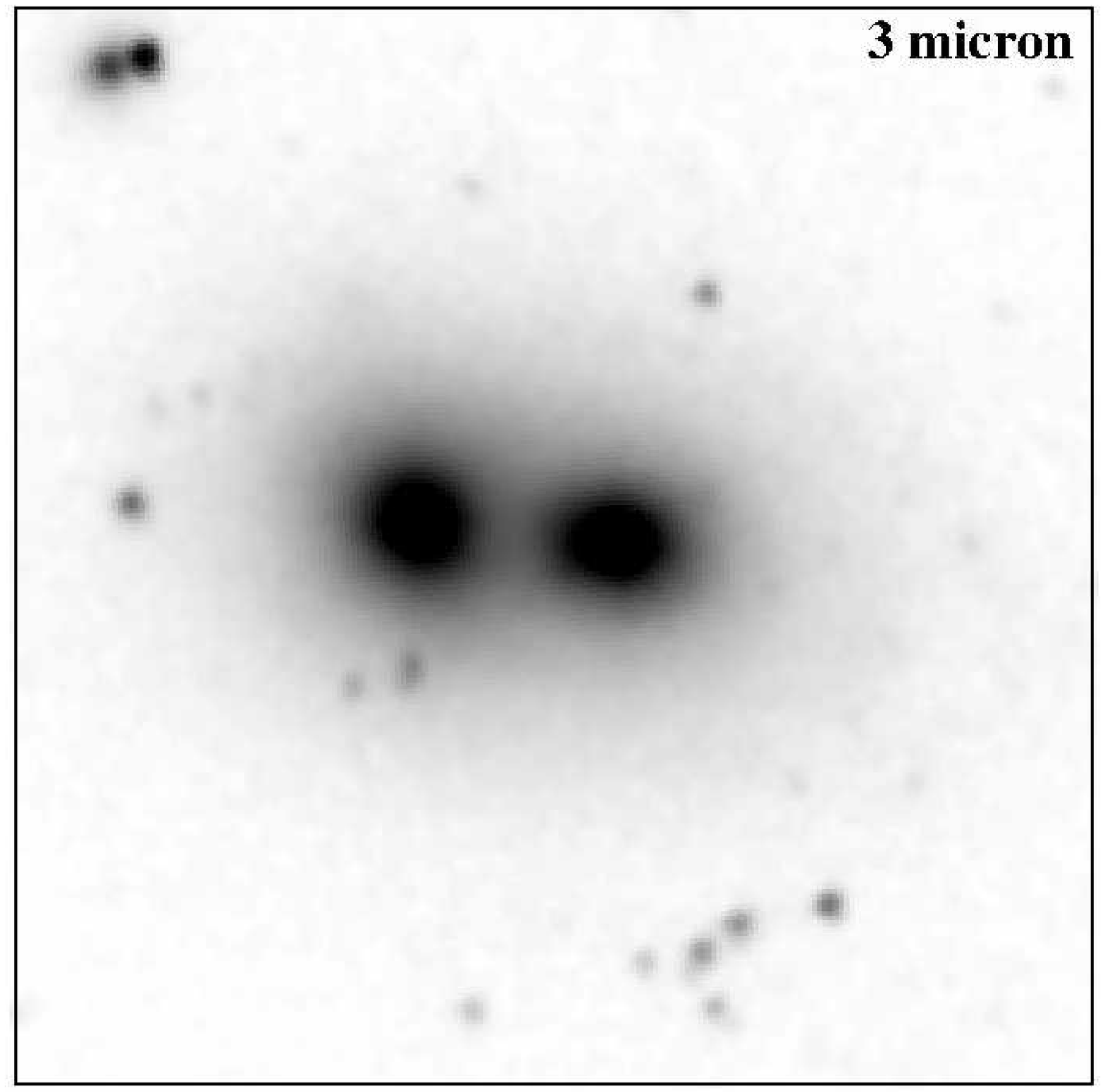}
\includegraphics[width=0.32\columnwidth]{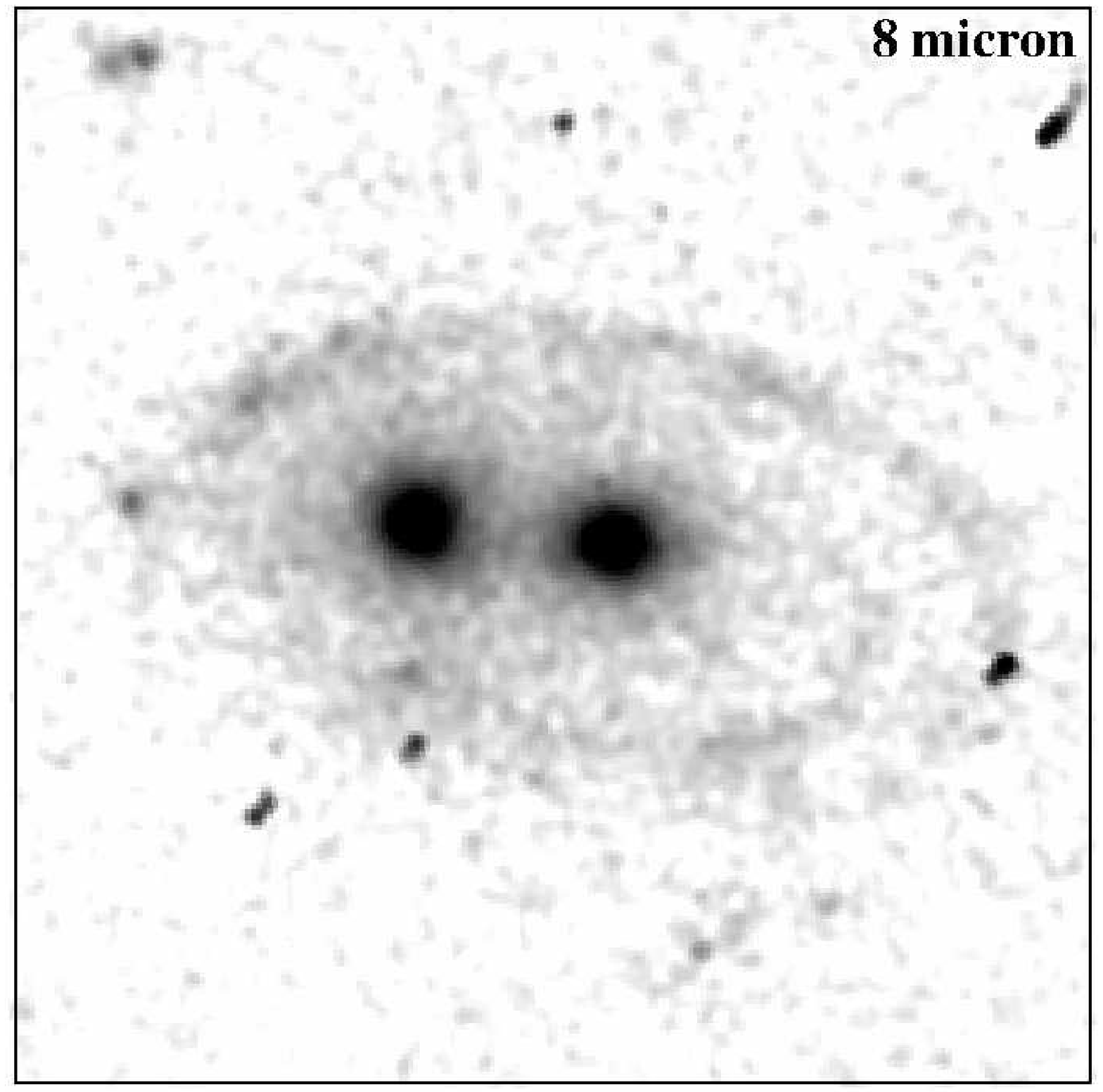}
\includegraphics[width=0.32\columnwidth]{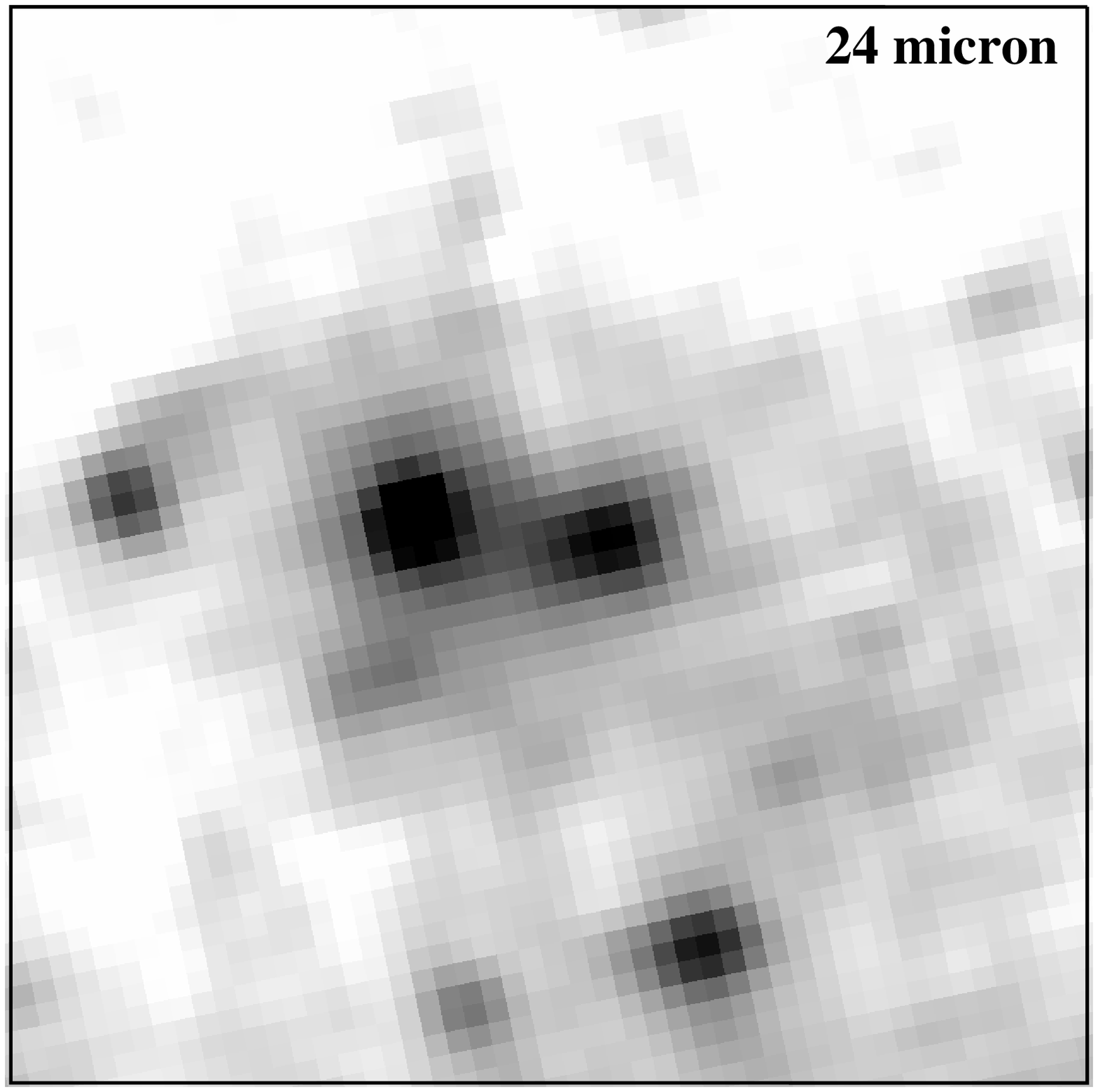}
\includegraphics[width=0.32\columnwidth]{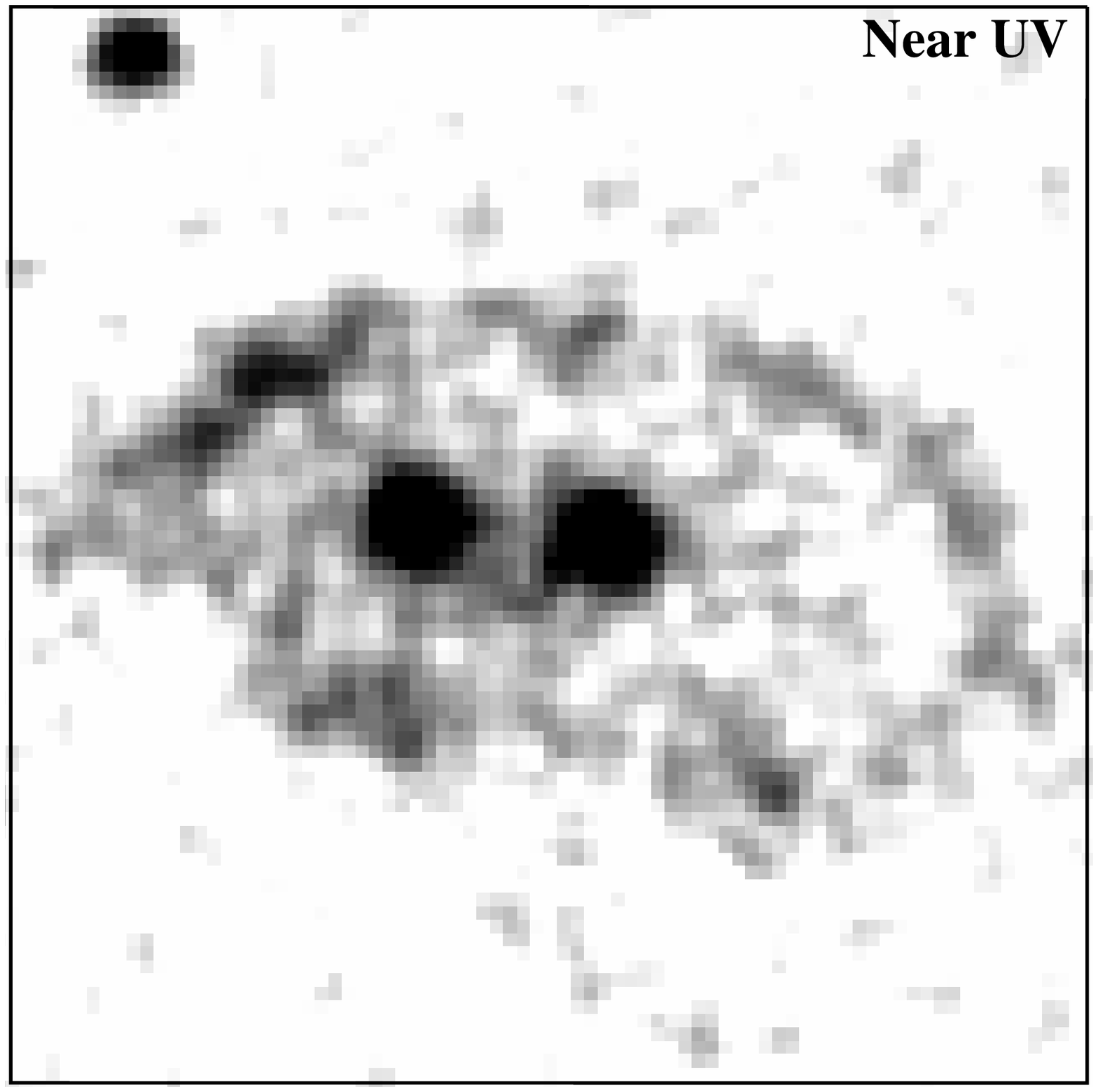}
\includegraphics[width=0.32\columnwidth]{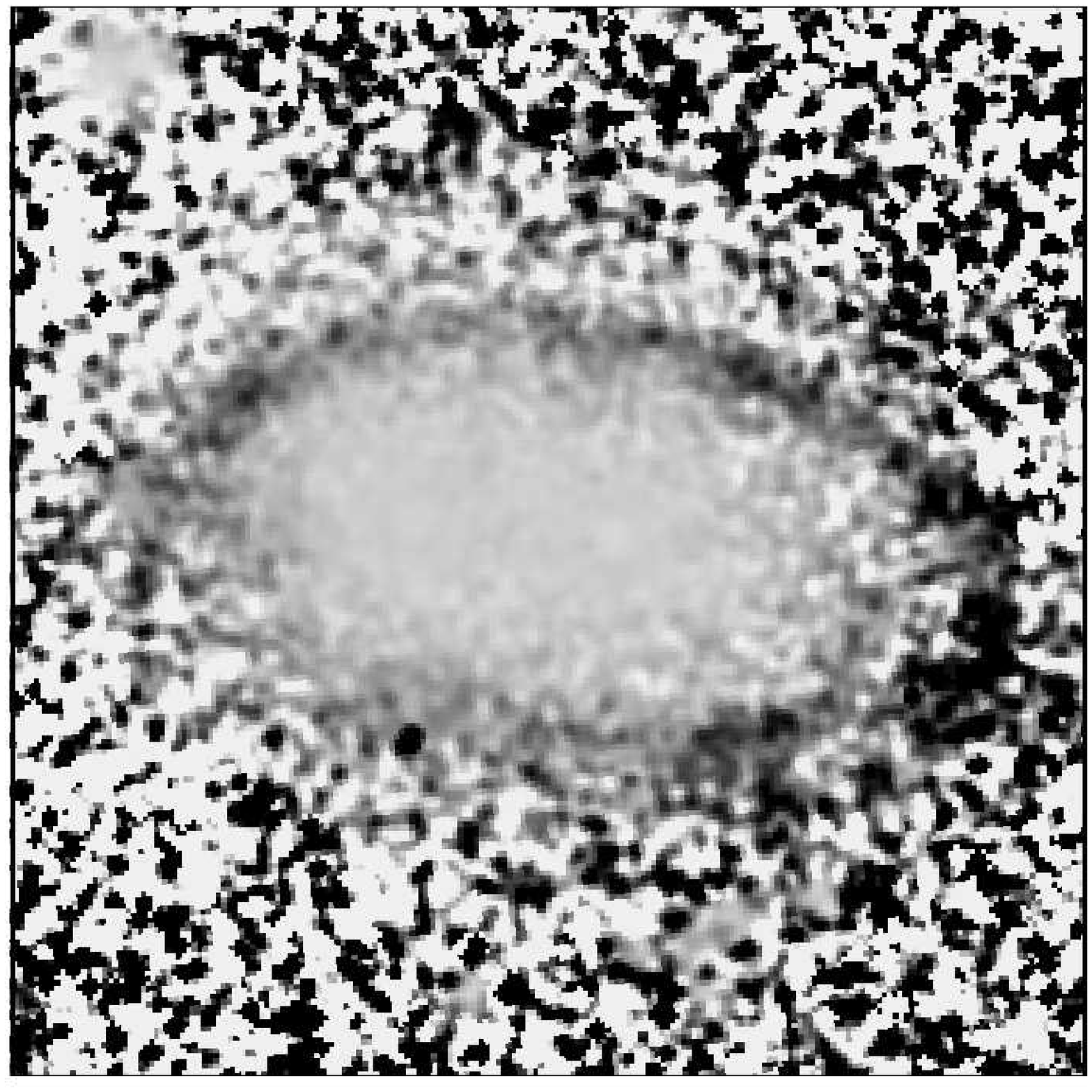}
\caption{\textit{Chandra} exposure-corrected image smoothed with a 2D Gaussian $\sigma=6\asec$ showing a $2\amin\times2\amin$ region around the NE group core (upper left).  The positions of the two elliptical galaxies are shown by the crosses.  \textit{Spitzer} IR images from IRAC at $3.6\mum$ (upper centre) and $8\mum$ (upper right) and MIPS at $24\mum$ (lower left) of the two central galaxies in the NE group (\citealt{Quillen08}).  The GALEX near-UV image is shown lower centre.  The sky-subtracted IRAC $3.6\mum$ image was also divided by the $8\mum$ image to show the ring emission more clearly (lower right).  The \textit{Spitzer} and GALEX images were smoothed with a 2D Gaussian $\sigma=1\asec$.  All images cover the same field of view.}
\label{fig:zoomin}
\end{minipage}
\end{figure*}

RXJ0751.3+5012 was observed with the \textit{Chandra} ACIS-I detector
for a single $100\ks$ exposure on 14 May 2013 (ID 15170).  There is
also an archival ACIS-I observation of $5\ks$ taken on 8 January 2011
(ID 12811) but this contributed few extended source counts and was therefore
not included in the analysis.  The new observation was
reprocessed with CIAO 4.5 and CALDB 4.5.9 provided by the
\textit{Chandra} X-ray Center (CXC).  The level 1 event file was
reprocessed to apply the latest gain and charge transfer inefficiency
correction and then filtered to remove any photons detected with bad
grades.  The observation was telemetred in VFAINT mode which provided
improved background screening.  A background light curve was extracted
from a source free region on chip 0 and filtered with the
\textsc{lc\_clean}\footnote{See
  http://cxc.harvard.edu/contrib/maxim/acisbg/} script to identify
periods in the observation affected by flares.  There was a short
flare towards the end of the observation reducing the final cleaned
exposure to $79.8\ks$.


Fig. \ref{fig:sbimage} (left) shows an exposure-corrected image of
RXJ0751.3+5012 covering the $0.3-4\keV$ energy band.  The raw image
was corrected for exposure variation by dividing by a
spectrum-weighted exposure map.  Point sources were identified with
\textsc{wavdetect} (\citealt{Freeman02}), confirmed visually and
excluded from the analysis.  The background was subtracted from images
and spectra using the appropriate blank-sky background observations.
Standard blank-sky backgrounds were processed identically to the
events file, reprojected to the corresponding sky position and
normalized to match the count rate in the $9.5-12\keV$ energy range.
The normalized blank-sky background spectrum for a source-free region
of ACIS-S2 was compared to the observed background spectrum and found
to be a close match over the whole energy range.


\begin{figure}
\centering
\includegraphics[width=0.95\columnwidth]{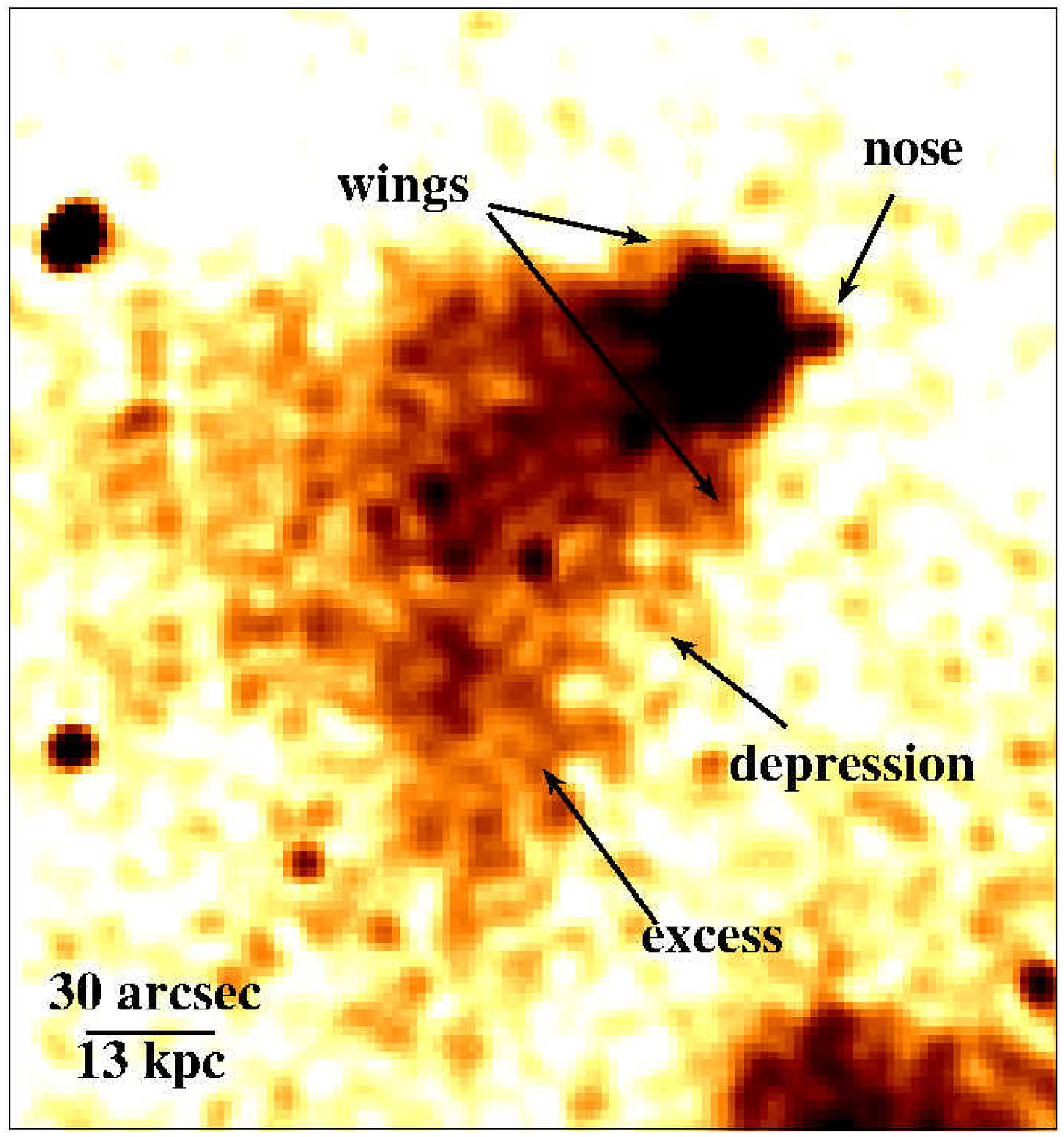}
\caption{Exposure-corrected image showing structure in the ram pressure stripped tail of the NE group.  The image covers the $0.3-4.0\keV$ energy band and has been binned by a factor of 4 and smoothed with a 2D Gaussian $\sigma=1.5\asec$.}
\label{fig:tailimage}
\end{figure}

\subsection{Imaging analysis}
\label{sec:imaging}

The \textit{Chandra} observation of RXJ0751.3+5012 reveals an off-axis
collision between two galaxy groups, where the north-eastern (NE)
group has recently passed to the north of the primary, south-western
(SW) group (Fig. \ref{fig:sbimage} left).  Ram pressure generated by
the motion through the ambient medium is stripping material from the
bright, dense core in the NE group.  The tail of stripped material
extends $\sim100\kpc$ ($3.5\amin$) and there is a trail of soft X-ray
gas blobs $\sim30\kpc$ behind the core that appears to curve towards
the SW group.  There is a sharp drop in surface brightness around the
N edge of the NE group core and this is likely to be a cold front
marking the leading edge of the core.  There is potentially a second,
outer surface brightness edge $\sim15\kpc$ ($30\asec$) ahead of the NE
group core (shown by arrows).  The SW group appears more diffuse with
a longer ($\sim190\asec$) surface brightness drop around its SE and N
edges.  This edge has a `boxy' morphology similar to that found in
A496 (\citealt{Dupke03,Dupke07}).  This edge is likely disrupted by
the passage of the NE group and interacting with the ram pressure
stripped tail, which could explain the appearance of substructure
around this interface.

Fig. \ref{fig:xmmimage} shows the merged MOS and PN image from the
$12\ks$ archival XMM-Newton observation (obs. ID 0151270201).  The
broad structure is consistent with two clear surface brightness peaks
associated with the two groups and surface brightness edges around
each core.  This image also reveals the larger scale hot atmosphere
that the two group cores are moving through and clear diffuse emission
ahead of the NE group.  The ram pressure generated by the
  motion of the core through the hot ambient medium is sufficient to
  strip the outer gas layers from the NE group producing the observed
  long tail of debris.  Similar structures have been observed behind
  dense, cool cores in major cluster mergers and in the Virgo cluster,
  where long plumes of gas extend behind infalling galaxies that are
  strongly interacting with the intracluster medium
  (eg. \citealt{Forman79}; \citealt{Nulsen82}; \citealt{Randall08M86};
  \citealt{Sun10}; \citealt{Million10}; \citealt{Kraft11};
  \citealt{Russell12}).


The DSS optical image (Fig. \ref{fig:sbimage} right) shows the
distribution of galaxies associated with the two groups.  The NE group
core hosts a pair of large elliptical galaxies (UGC4052) separated by
a projected distance of only $15\kpc$ (\citealt{Crawford99}).  The
eastern galaxy of the pair, UGC4052 E, is coincident with the X-ray
surface brightness peak in the core and hosts luminous H$\alpha$ line
emission ($L_{\mathrm{H}\alpha}\sim10^{40}\ergps$,
\citealt{Crawford99}).  UGC4052 W is located ahead of the NE group
core and appears associated with a small spur of X-ray emission or
`nose' extending $\sim7\kpc$ ($15\asec$; Fig. \ref{fig:zoomin} upper
left).  Another large elliptical galaxy (UGC4051) is coincident with
the X-ray surface brightness peak in the SW group.  The redshifts for
these three large ellipticals from \citet{Crawford99} show that these
galaxies are associated with the two merging groups and not a chance
superposition along the line of sight. There is a hard X-ray point
source spatially coincident with UGC4051 in the SW group, which could
indicate an AGN.  The VLA FIRST survey does not show a radio
point source coincident with this source (\citealt{Becker95}).  There is a radio point
source spatially coincident with UGC4052 E in the NE group.
Although there does not appear to be a hard X-ray point source
associated with the radio point source, this could be simply too faint
to detect in the existing data.

The ram pressure stripped tail behind the NE group core shows
irregular structures or `wings' (Fig. \ref{fig:tailimage};
\citealt{Machacek06}; \citealt{Kraft11};
\citealt{Roediger12Groups,Roediger12}).  The northern wing
  contains $\sim110$ counts and the southern wing contains $\sim190$
  counts, which corresponds to $>10\sigma$ above the surrounding
  ambient level for each structure.  Distortions in cold fronts can be
  generated by the shear flow around them, which produces
  Kelvin-Helmholtz instabilities.  The shear flow is expected to be
  maximum along the sides of the NE group tail, which explains the
  contrast between the narrow N edge of the cold front and the
  distorted structure at larger angles to the direction of motion.
  Non-viscous hydrodynamical simulations of cold fronts in the
  intracluster medium commonly show the development of these
  instabilities (eg. \citealt{Murray93}; \citealt{Heinz03};
  \citealt{Zuhone10}; \citealt{Roediger11}) and can reproduce the
  structure of distortions in observed cold fronts
  (eg. \citealt{Roediger12, Roediger12Groups}).  There is also an
  arc-like surface brightness excess and depression indicative of a
  hydrodynamical instability along the S side of the tail (see section
  \ref{sec:radprofiles}, eg. \citealt{Mazzotta02};
  \citealt{Roediger13}).  

The `nose' ahead of the NE group contains $\sim100$ counts and is
detected at more than $5\sigma$ above the ambient medium.  This structure is
more likely due to ram pressure stripping of an X-ray corona than a
turbulent distortion of the cold front (Fig. \ref{fig:zoomin} upper
left; \citealt{Sun05}; \citealt{Sun05Perseus}).  The N surface
brightness edge appears very narrow and smooth on either side of this
structure, rather than turbulent, and the `nose' is spatially
coincidence with a large elliptical galaxy, which is a likely host for
an X-ray corona (\citealt{Forman85}; \citealt{Vikhlinin01Coma};
\citealt{Sun07}).  We discuss these structures in more detail in
Sections \ref{sec:core} and
\ref{sec:tail}.

UGC4052 was included in the \citet{Quillen08} \textit{Spitzer} study
of the mid-IR emission from the central galaxies in groups and
clusters with optical emission lines.  \citet{Quillen08} found that
UGC4052 E has excess IR emission above that expected for a quiescent
elliptical galaxy, which is likely due to hot dust powered by star
formation.  \citet{ODea08} estimated a star formation rate for this
galaxy from the IR luminosity of $0.7\Msunpyr$.  The \textit{Spitzer}
$8\mum$ image (Fig. \ref{fig:zoomin} upper right) shows a striking
bright oval ring of emission surrounding both galaxies with a diameter
of $\sim40\kpc$ ($90\arcsec$).  This feature is seen primarily at
$8\mum$, which is likely due to a polycyclic aromatic
  hydrocarbon (PAH) emission feature at $7.7\mum$, but is also seen
faintly at $24\mum$ and in \textit{GALEX} near-UV observations
(Fig. \ref{fig:zoomin} lower left and centre).  The ring appears
broken in places to the east and west and contains several knots of
enhanced emission.  

\subsection{Radial profiles}
\label{sec:radprofiles}

\begin{figure*}
\begin{minipage}{\textwidth}
\centering
\includegraphics[width=0.48\columnwidth]{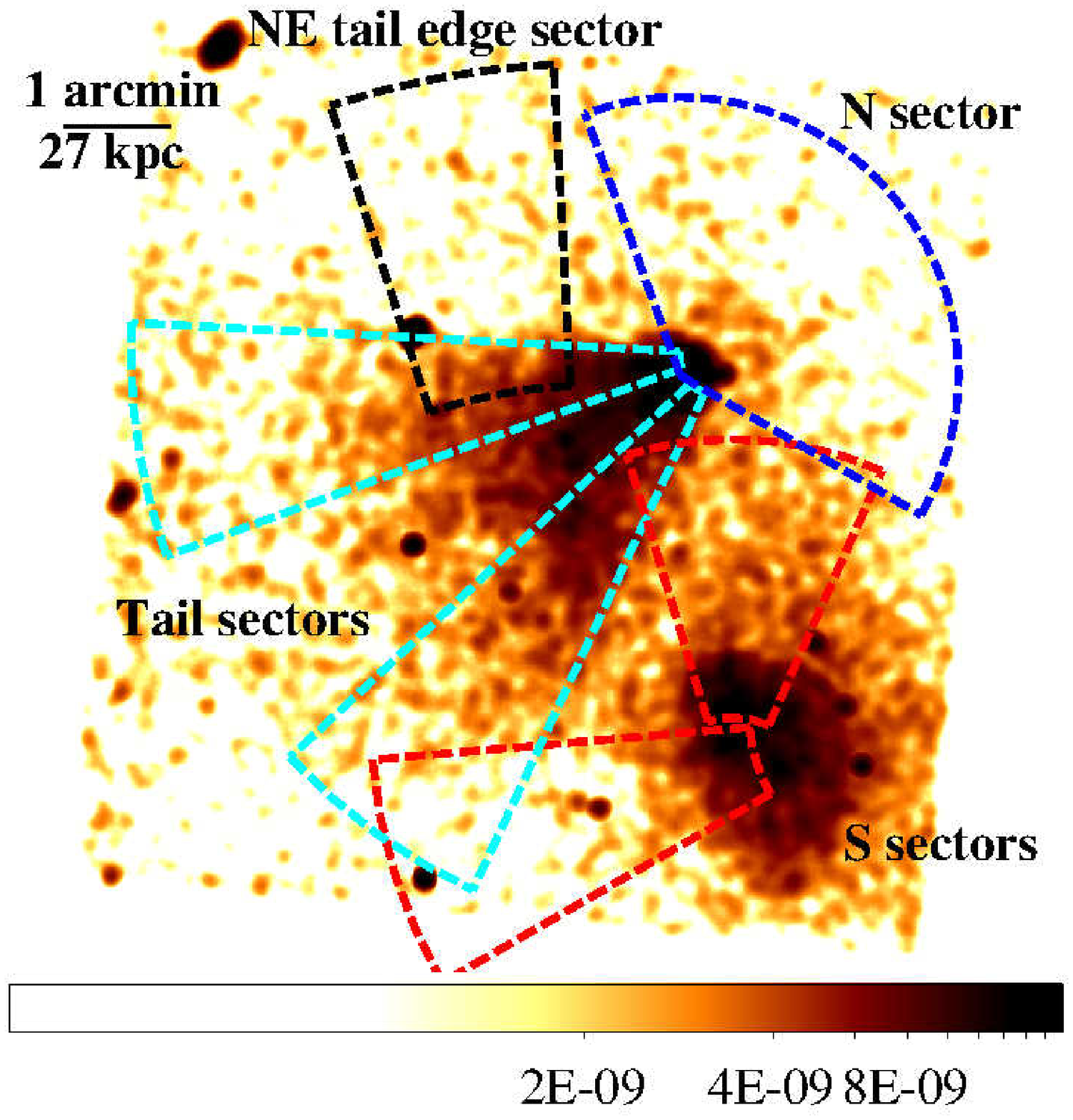}
\includegraphics[width=0.48\columnwidth]{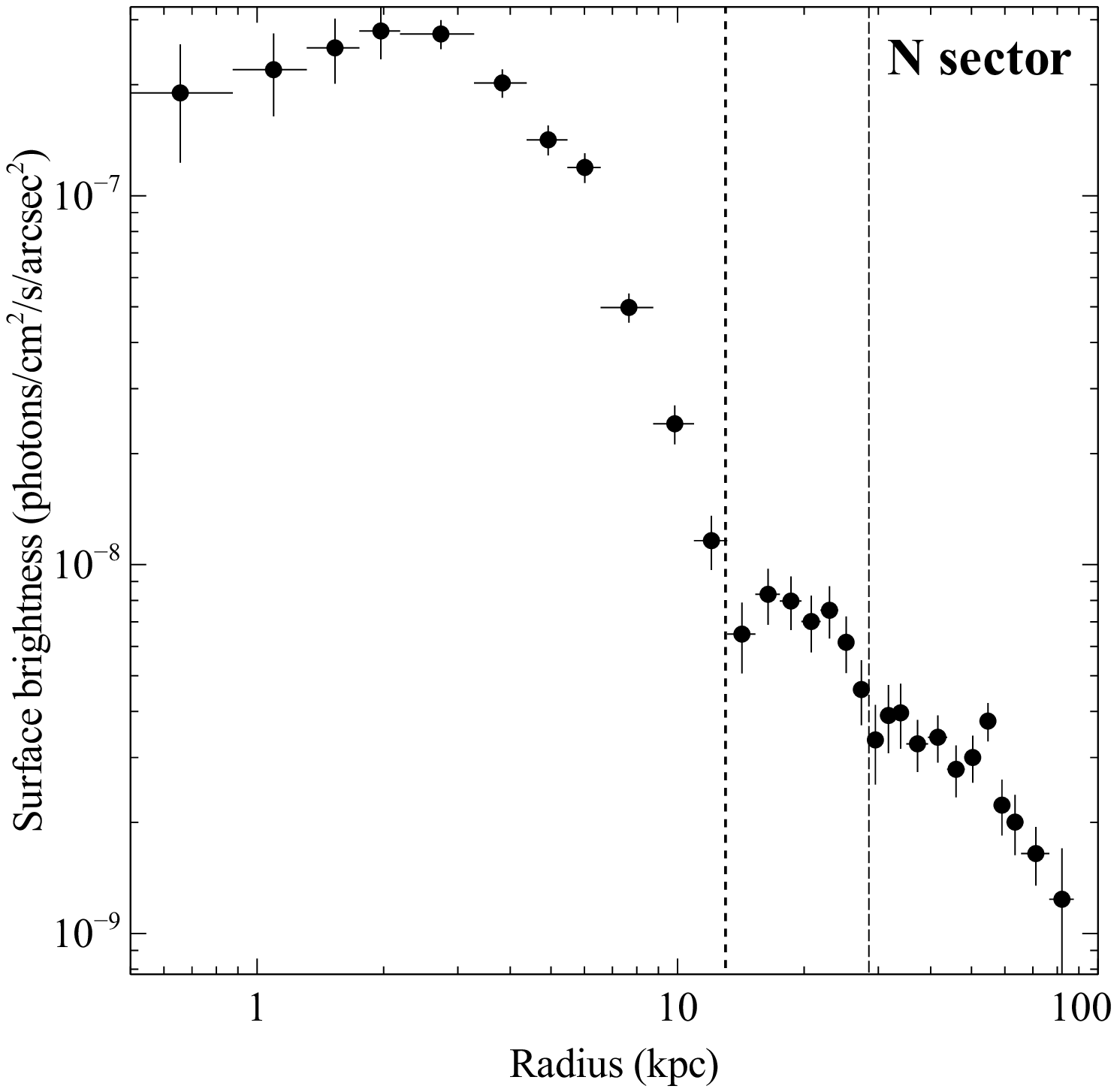}
\includegraphics[width=0.48\columnwidth]{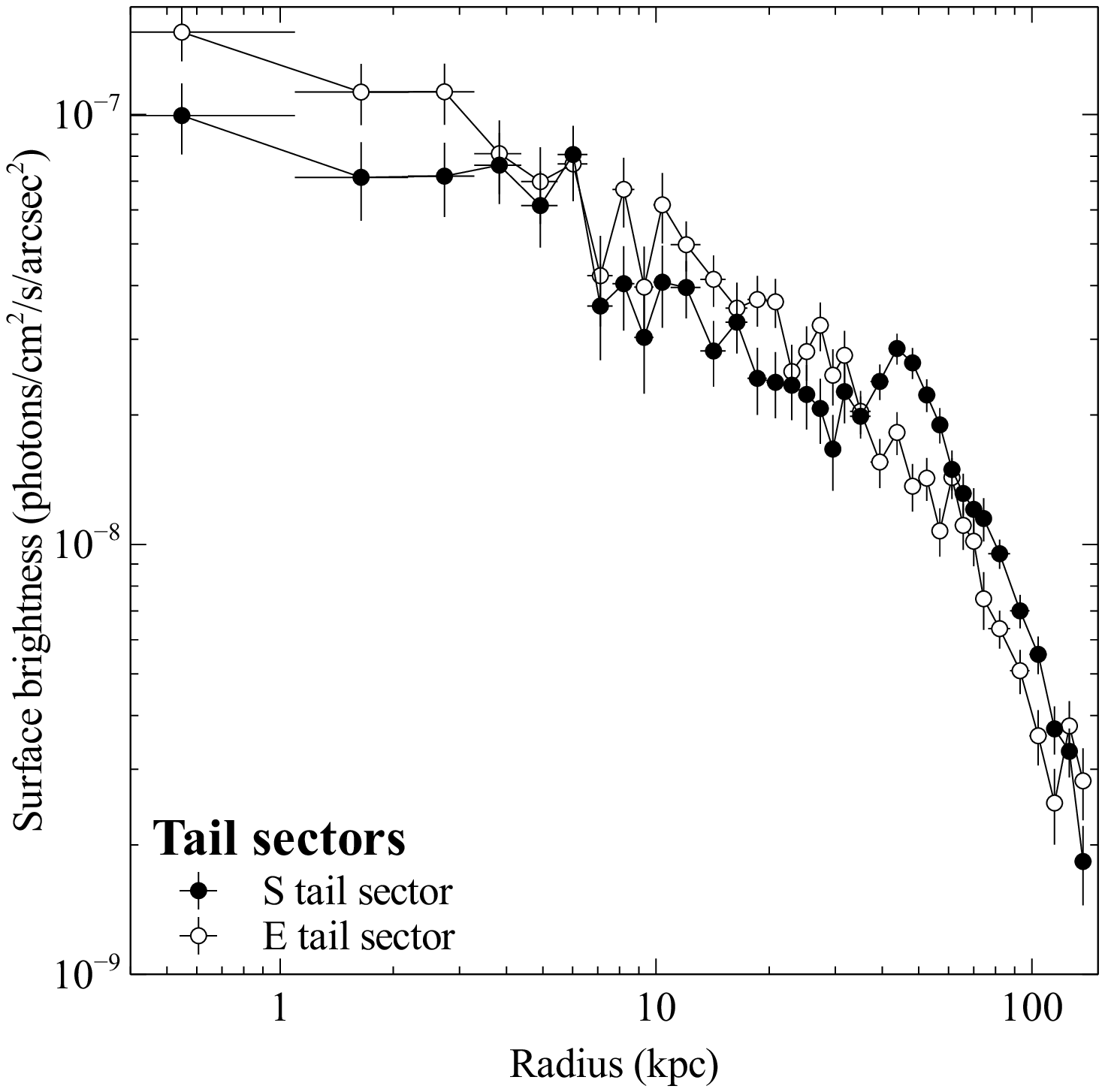}
\includegraphics[width=0.48\columnwidth]{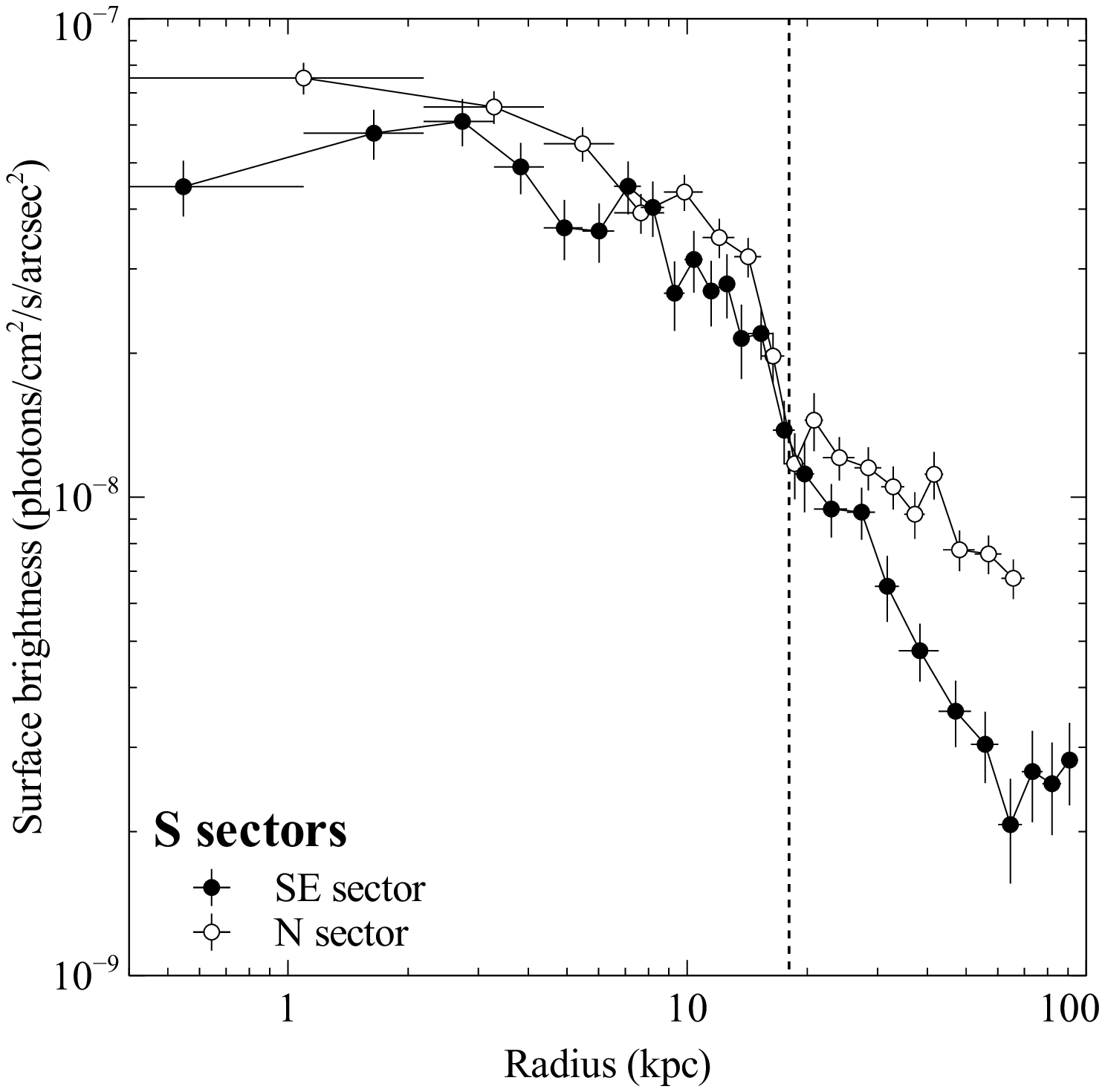}
\caption{Background-subtracted surface brightness profiles in the
  energy band $0.3-4.0\keV$ for the N sector, Tail sectors and S sectors shown in the
  exposure-corrected image (upper left).  Upper right: N sector from
  the centre of the NE subcluster across the leading edge.  Lower left:
  profiles for sectors through the NE group subcluster tail.  Lower right: profiles for the S sectors in the SW group.  The surface brightness edges are marked by dotted and dashed lines.  The surface brightness profile for the NE tail edge sector is shown in Fig. \ref{fig:tailSBNedge}.}
\label{fig:SBsectors}
\end{minipage}
\end{figure*}

We extracted projected surface brightness profiles across the edges
identified in Section \ref{sec:imaging} and through the ram pressure
stripped NE group tail.  Fig. \ref{fig:SBsectors} (upper left) shows
the regions selected, where the centre of each sector was selected to
coincide with the centre of curvature of the surface brightness edges.
Radial bins were $1\arcsec$ in width for the bright, core regions and
increased in width at larger radii to ensure a minimum of 30 source
counts per spatial bin.  The surface brightness profile for the N
sector (Fig. \ref{fig:SBsectors} upper right) shows the clear N edge at
$14\kpc$ where the surface brightness drops by a factor of $\sim10$
over a radial distance of $\sim7\kpc$ across the leading edge of the
NE group core.  The N sector includes the `nose' structure, which
broadens the edge.  The outer surface brightness edge tentatively
identified in Fig. \ref{fig:sbimage} is clearly detected as a drop by
a factor of $\sim2$ at a radius of $30\kpc$.  The location of this
edge ahead of the NE group core suggests it could be a bow shock.  In
Section \ref{sec:spatialspec}, we extract the temperature profiles
across these edges to determine if there is a corresponding
temperature jump.

Fig. \ref{fig:SBsectors} (lower left) shows surface brightness
profiles through the ram pressure stripped tail of the NE group.  The
S and E tail sectors show a similar decline in surface brightness with
radius.  However, there is a clear excess in the S tail sector from
$30\kpc$ to $60\kpc$ radius, which corresponds to the excess surface
brightness identified in Fig. \ref{fig:tailimage}.  This surface
brightness excess is preceded by a region of reduced surface
brightness, by comparison with the E sector, from $10\kpc$ to $30\kpc$
radius.  The profiles confirm that these are significant structures
and could correspond to a dynamical instability resulting from the
merger.  The location of this feature appears similar to that of the
large instability identified in the merging cluster A3667
(\citealt{Mazzotta02}; see also \citealt{Roediger13}) and we discuss
this further in Section \ref{sec:tail}.

The surface brightness profiles across the edge around the SW group
core are shown in Fig. \ref{fig:SBsectors} (lower right).  The surface
brightness drops by a factor of $\sim2$ at a radius of $20\kpc$ in
both of the sectors analysed and the edge appears consistently narrow
for $\sim90\kpc$ around the N and SE sides of the core.  However, this analysis
was limited due to the overlap with the ram pressure stripped material
from the NE group core and the low number of counts did not allow us to
place a strong upper limit on the width.

\subsection{Spatially resolved spectroscopy}
\label{sec:spatialspec}


The total spectrum for each group was extracted from a circular region
centred on each X-ray peak with a radius of $1\arcmin$, which
contained the bulk of the emission and excluded point
sources.  A corresponding background spectrum was extracted from the
normalized blank sky dataset and appropriate responses and ancillary
responses were generated.  Each spectrum was restricted to the energy
range $0.5-7\keV$ with a minimum of 20 counts per spectral channel.
The spectra were fitted in \textsc{xspec} 12.8 (\citealt{Arnaud96})
with an absorbed single temperature \textsc{apec} model version 2.0.2
(\citealt{Smith01}).  The cluster redshift was fixed to $z=0.022$
(\citealt{Ebeling98}; \citealt{Crawford99}) and abundances were
measured assuming the abundance ratios of \citet{AndersGrevesse89}.  The temperature, metallicity and normalization parameters were left free.

The absorbing column density was determined using the spectrum extracted
from the region covering the SW group, which excluded
the cooler gas components in the NE group core and tail.  The
absorption parameter was left free and the best-fit value was
$N_{\mathrm H}=0.08\pm0.02\times10^{22}\psqcm$.  This is a little
higher than the Galactic absorption found in HI surveys
(\citealt{Kalberla05}) but a similar analysis for the earlier Chandra
observation also indicates absorption above the Galactic value.  We
therefore fixed the absorption parameter to this best-fit value for
all other spectral fits.

\begin{figure}
\centering
\includegraphics[width=0.95\columnwidth]{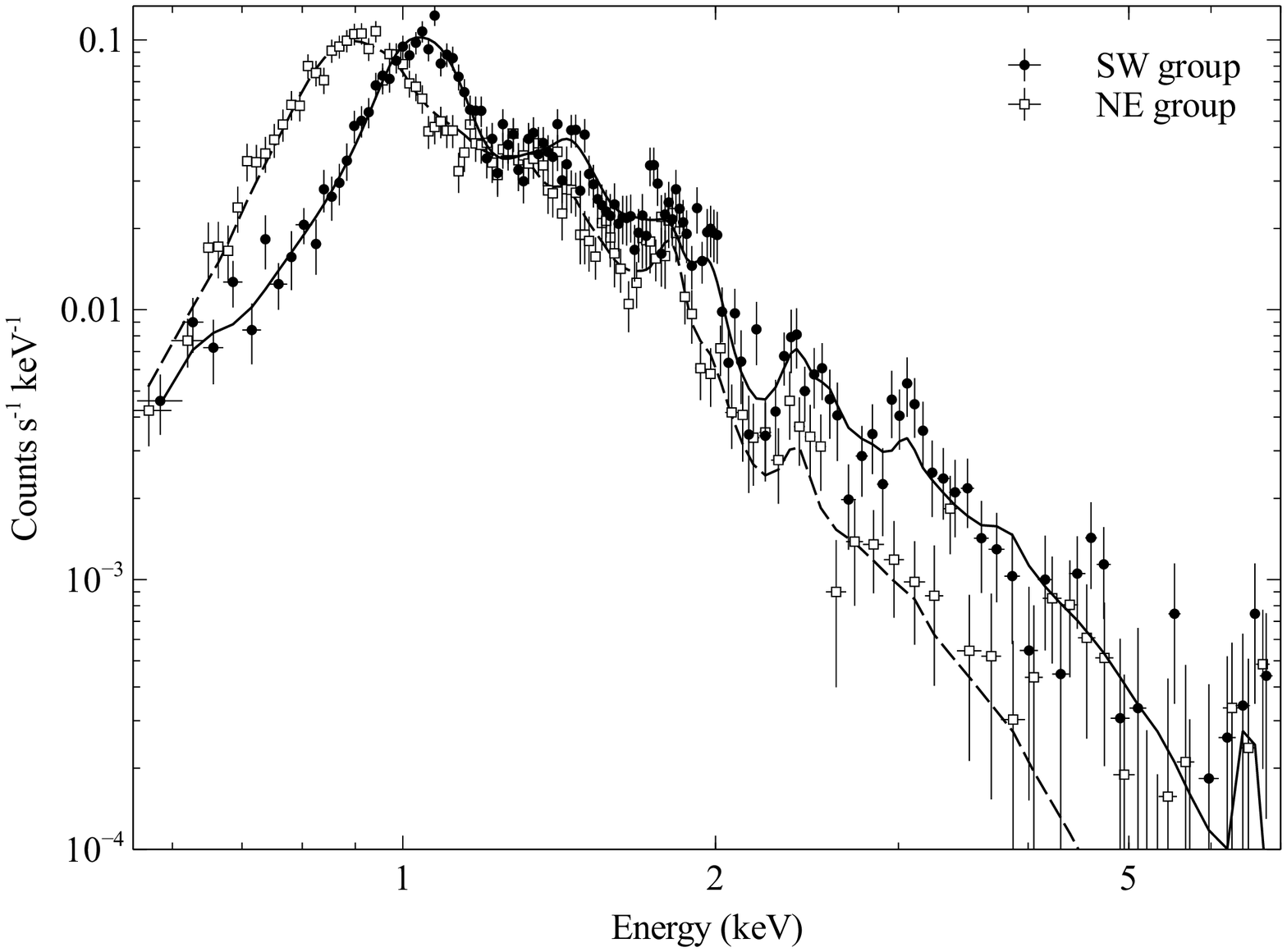}
\caption{The total background subtracted spectrum for the NE group (squares) and the SW
  group (circles) and the corresponding best-fit absorbed
  \textsc{apec} models (solid lines).}
\label{fig:globalspec}
\end{figure}

The total spectrum for each group is quite different
(Fig. \ref{fig:globalspec}).  The SW group has a global temperature of
$1.60^{+0.02}_{-0.03}\keV$ and a metallicity of
$0.71^{+0.10}_{-0.09}\Zsun$.  The NE group is much cooler and
metal-poor with a temperature of $0.84\pm0.01\keV$ and a metallicity
of $0.16\pm0.01\Zsun$.  We also used a two temperature spectral model
to determine if the lower metallicity in the NE group was a result of
the Fe bias at low temperatures (\citealt{Buote00}).  However, the
single temperature model was found to be a good fit to both total
spectra and we did not find any significant additional temperature
component in either galaxy group.  This observed difference in
metallicity follows the trend of higher metal abundances in the
central regions of higher temperature groups
(eg. \citealt{Rasmussen09}).  We repeated the single temperature
spectral fits leaving the redshift parameter free and obtained values
of $z=0.022^{+0.001}_{-0.002}$ and $z=0.026^{+0.001}_{-0.003}$ for the
SW and the NE groups, respectively.  Although this is not a
significant difference, these results are consistent with the galaxy
redshifts $z=0.0218$ for UGC4051 in the SW group and $z=0.0236$ and
$z=0.0228$ for UGC4052 E and W, respectively (\citealt{Crawford99}).
From the global temperature of each group and the mass-temperature
relation (eg. \citealt{Finoguenov01}; \citealt{Sanderson03};
\citealt{Vikhlinin06}), we estimate that the SW group is roughly three
times more massive than the NE group.  The global temperature of the
NE group is likely to have been underestimated as ram pressure
stripping has removed most of the warmer, lower density gas.  However, adiabatic compression due to the ram pressure and static pressure of the larger group will counteract this to some extent.  The mass
ratio is therefore likely to be an upper limit.

Maps of the projected gas properties were produced using the Contour
Binning algorithm (\citealt{Sanders06}), which generates spatial bins
that closely follow the X-ray surface brightness.  For the
temperature, normalization and pressure maps, regions were selected to
provide a signal-to-noise ratio of 22 ($\sim480$ counts).  Larger
regions with a signal-to-noise ratio of 32 were used for the
metallicity map.  The maps are $8\amin\times8\amin$ covering the two
groups.  The length of the spatial regions was constrained to be at
most twice their width.  For each region, a spectrum was extracted
from the Chandra dataset and restricted to the energy range
$0.5-7\keV$.  A corresponding background spectrum was extracted from
the normalized blank sky dataset and appropriate responses and
ancillary responses were generated.  An absorbed single temperature
\textsc{apec} model was fitted to each spectrum with the temperature
and normalization parameters left free and minimising the C-statistic
(\citealt{Cash79}).  The metallicity was fixed to $0.3\Zsun$ for the
temperature, normalization and pressure maps that were produced with
fewer counts per spatial region.

\begin{figure*}
\begin{minipage}{\textwidth}
\centering
\includegraphics[width=0.48\columnwidth]{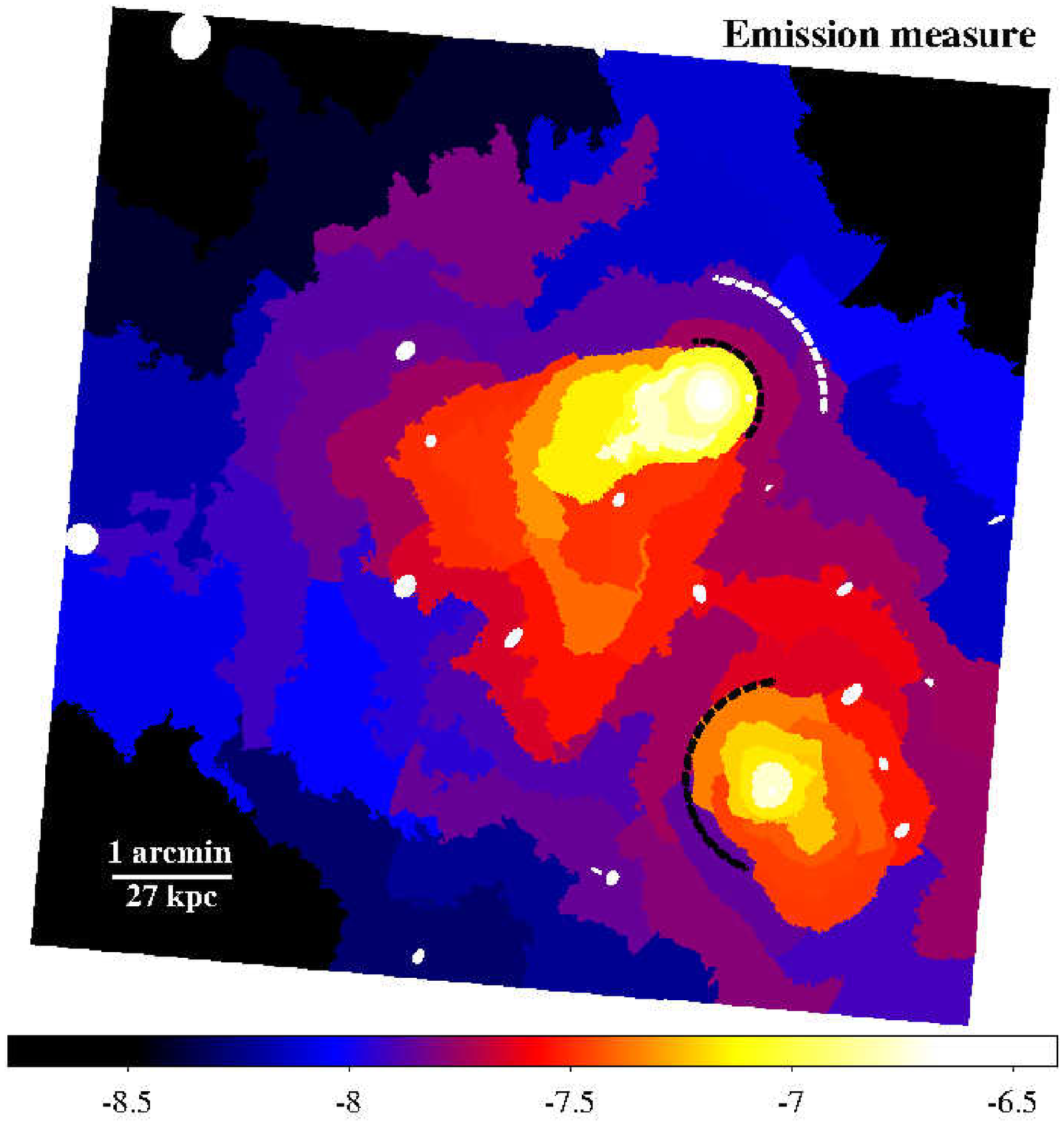}
\includegraphics[width=0.48\columnwidth]{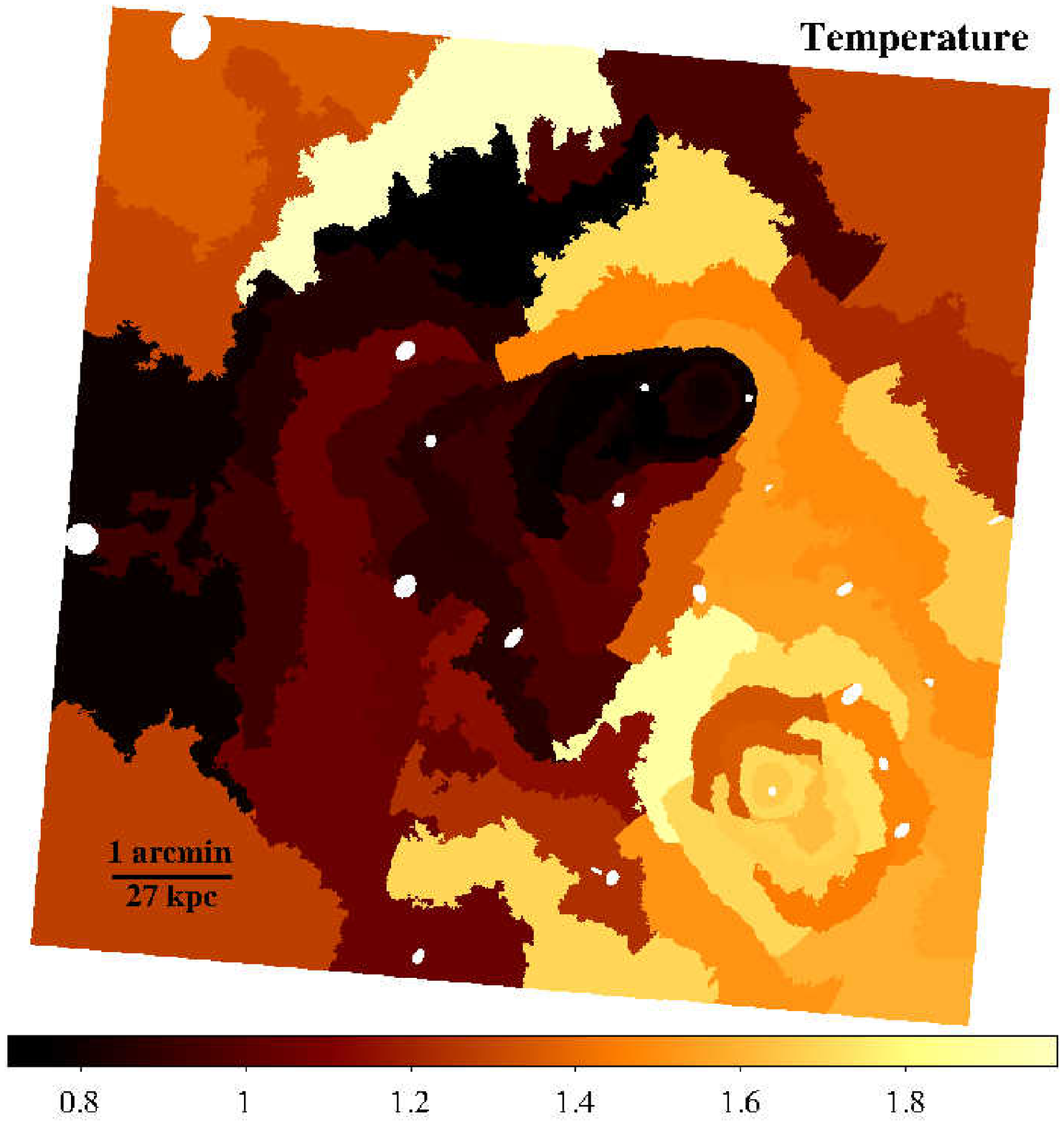}
\includegraphics[width=0.48\columnwidth]{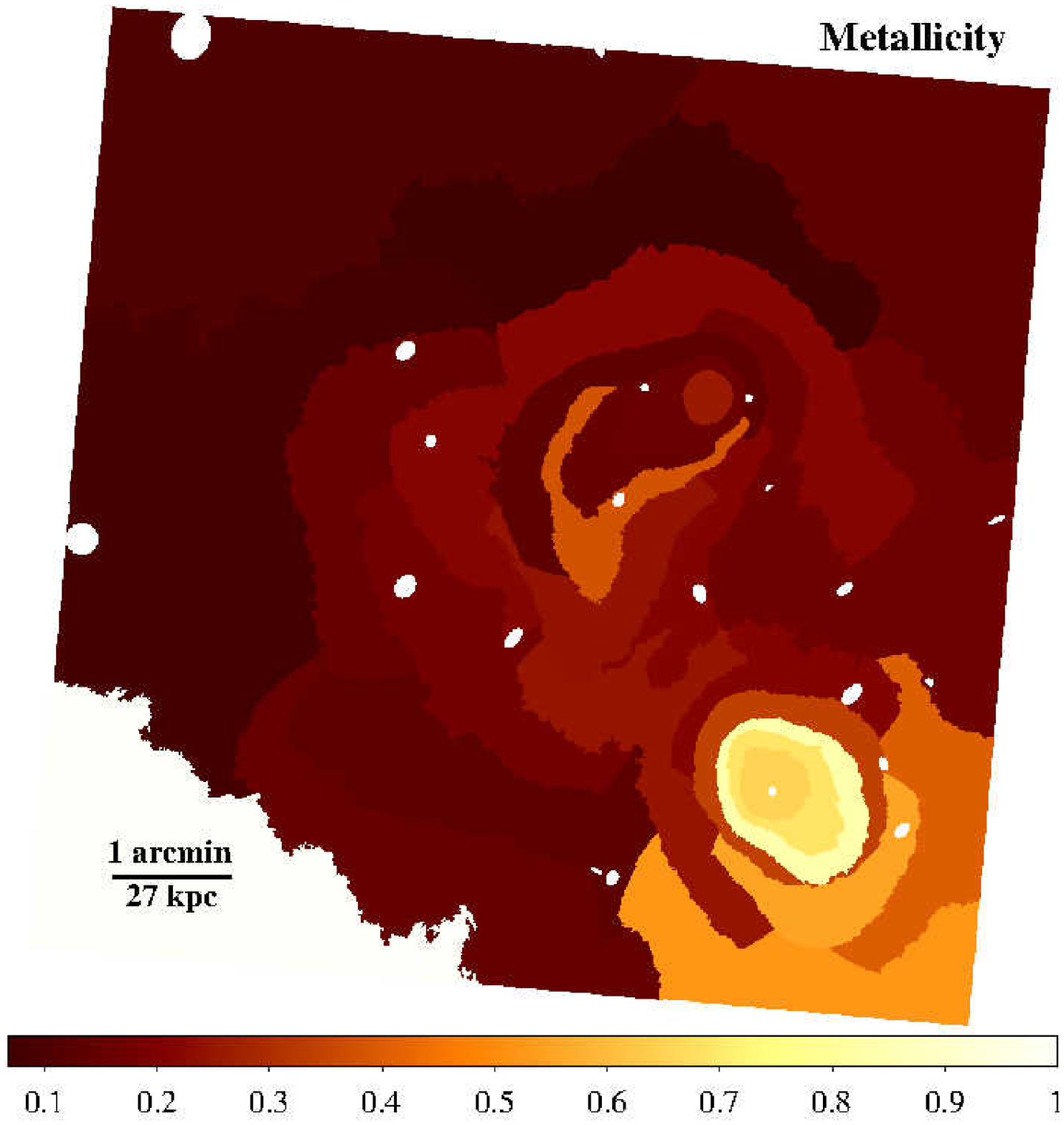}
\includegraphics[width=0.48\columnwidth]{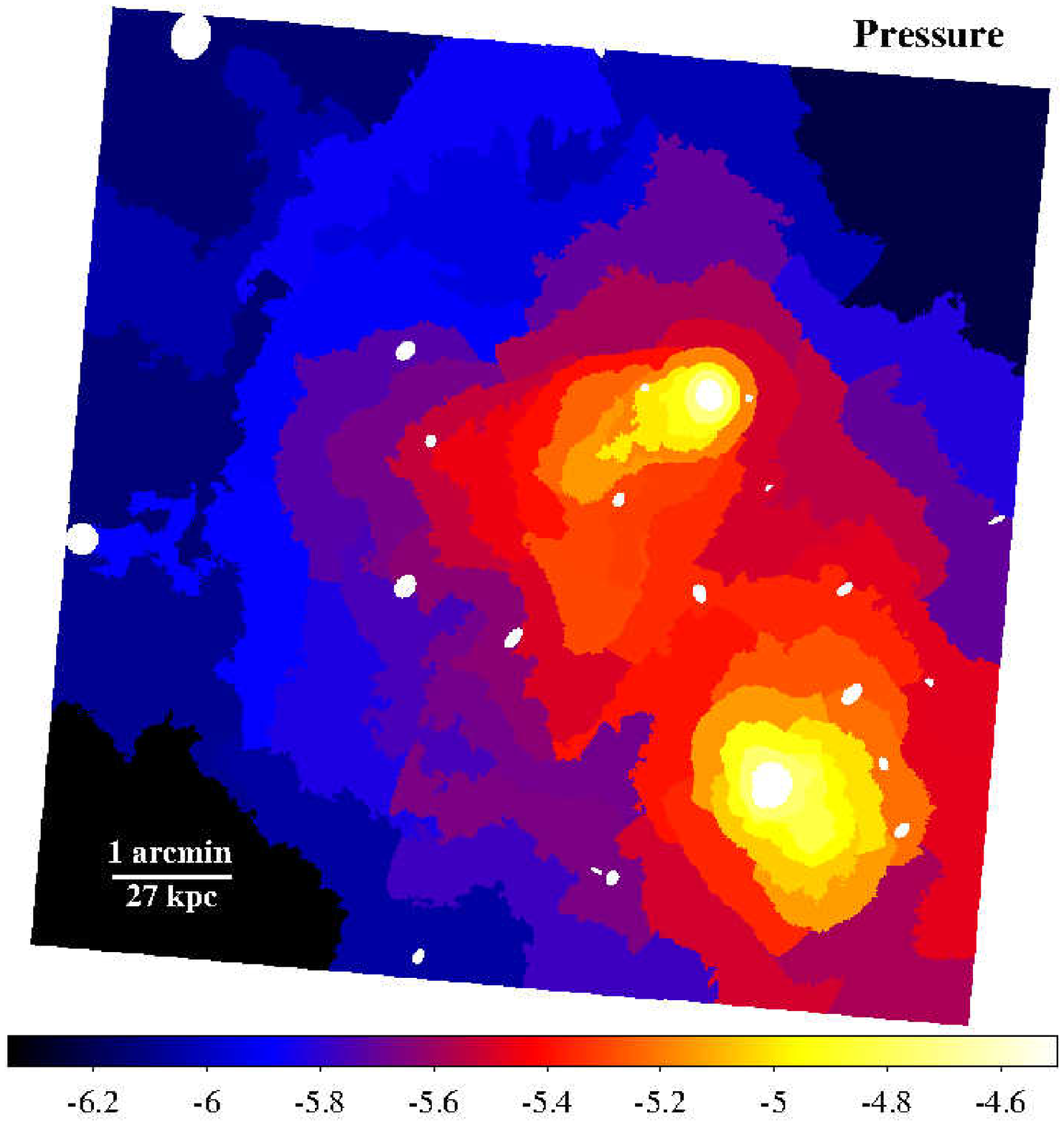}
\caption{Upper left: projected emission measure per unit area map (units are $\mathrm{log_{10}}\empasecsq$).  The emission measure is the
  \textsc{xspec} normalization of the \textsc{apec} spectrum
  $K=EI/(4\times10^{14}\pi D_A^2(1+z)^2)$, where EI is the emission
  integral $EI=\int n_en_H\mathrm{d}V$.  Upper right: projected temperature map (units keV).  Lower left: projected metallicity map ($\Zsun$) generated using larger spatial bins.  Lower right: projected pseudo-pressure map per unit area ($\mathrm{log_{10}}\pseudoP$).  The excluded point sources are visible as small white circles.  The positions of the cold fronts and shock front are shown on the emission measure map as black and white dashed lines, respectively.}
\label{fig:conbinmaps}
\end{minipage}
\end{figure*}

Fig. \ref{fig:conbinmaps} shows detailed maps of the best-fit values
for the projected emission measure per unit area, temperature and
metallicity.  The projected emission measure traces the square of the
gas density in the group.  The projected `pressure' map was produced
by multiplying the temperature map and the square root of the
normalization per unit area map.  The uncertainty in the normalization
and temperature values at the centre of each group are $\sim4\%$ and
$\sim6\%$, respectively.  The temperature is below $1.5\keV$ here and
Fe L line emission provides a good temperature diagnostic.  At large
radii, the uncertainty increases to $\sim10\%$ for the temperature and
$\sim7\%$ for the normalization.  The uncertainty in the metallicity
values is $\sim20\%$.


There is a clear peak in the emission measure at the centre of each of
the groups.  The surface brightness drops to the NW of the NE group
and along the SE edge of the SW group can be seen in the emission
measure map.  The emission measure declines smoothly through the tail
of ram pressure stripped material behind the NE group core.  The trail
of densest gas in the tail appears to bend towards the S approximately
$25\kpc$ ($55\asec$) behind the core.  At this point the tail also
appears to flare out rapidly, increasing in width from $20\kpc$
($40\asec$) to at least $70\kpc$ ($160\asec$).  There is a region of
increased emission measure to the N of the SW group core suggesting there
could be some ram pressure stripping.  However, the SW group core does
not appear strongly disrupted by the collision, which is on the
borderline between a major and minor merger.

The temperature map shows that the temperature in the NE group drops
below $0.8\keV$ in the dense core.  The S group is much hotter,
peaking at $1.71^{+0.09}_{-0.07}\keV$ and the temperature increases
across the surface brightness edge from $1.34^{+0.07}_{-0.06}\keV$ to
$1.7\pm0.1\keV$, indicative of a cold front.  The `boxy' morphology of
the cold front, similar to A496 (\citealt{Dupke03,Dupke07}), and the
lack of strong disruption to the SW group suggests that this is a
young sloshing cold front.  The merger has displaced the hot
atmosphere in the gravitational potential minimum and it is now
starting to oscillate, producing the observed cold fronts
(eg. \citealt{Markevitch07}).  \citet{Roediger12} found that the boxy
structure in A496 arises due to the development of Kelvin-Helmholtz
instabilities along cold fronts.  The cold front appears strongest
around the N edge of the SW group core where the outer hot atmospheres
of each group are colliding.  The entropy of the gas in this region
appears comparable to that located to the SE of the SW group core
suggesting that the gas has been compressed here by the collision
between the two groups.

There is also a cold front around the leading edge of the NE group
where the temperature increases much more steeply from
$0.79\pm0.04\keV$ to $1.5^{+0.2}_{-0.1}\keV$. The temperature
increases through the ram pressure stripped tail with the gas along
the S edge warmer than that along the NE edge.  The NE edge of the
tail appears much narrower, suggesting that there is much less mixing
with the hotter, ambient medium, consistent with the more gradual
increase in temperature.  The outer surface brightness edge ahead of
the NE group core appears coincident with a temperature drop from
$1.7^{+0.4}_{-0.3}\keV$ to $1.3\pm0.1\keV$.  This could therefore be a
bow shock and we analyse this structure in detail in Section
\ref{sec:shock}.


The SW group has a strong, central metallicity peak of
$0.8^{+0.3}_{-0.2}\Zsun$.  The NE group appears metal-poor by
comparison with a peak of only $0.35^{+0.14}_{-0.09}\Zsun$ in the
dense core.  The metallicity is lowest at $0.10^{+0.03}_{-0.02}\Zsun$
for $27\kpc$ ($60\asec$) behind the NE group core and then shows a
modest increase, particularly towards the S tail edge where there is likely
to be mixing with metal-rich material stripped from the SW group.  The
projected `pressure' map shows the extent of the shock
heating between the trajectory of the two groups.  There is also a
decrease in pressure ahead of the NE group core consistent with a bow
shock.


\subsection{Hardness Ratio}

Although there are too few counts to produce a spectrum for the NE
group `nose', we have produced a hardness ratio profile across this
region to look for evidence of temperature variations.
We have used the Bayesian method of calculating hardness ratios
detailed by \citet{Park06}, which is particularly useful for the low
counts regime.  The hardness ratio was calculated as the fractional
difference between the number of counts in the soft ($0.5-1.5\keV$)
and hard ($1.5-5.0\keV$) energy bands,

\begin{equation}
HR\equiv\frac{H-S}{H+S}.
\end{equation}

\noindent The source counts in each energy band were extracted from a
series of regions covering a narrow sector across the two galaxies.
The blank sky background counts were also modelled as independent
Poisson random variables.  

\begin{figure}
\centering
\includegraphics[width=0.8\columnwidth]{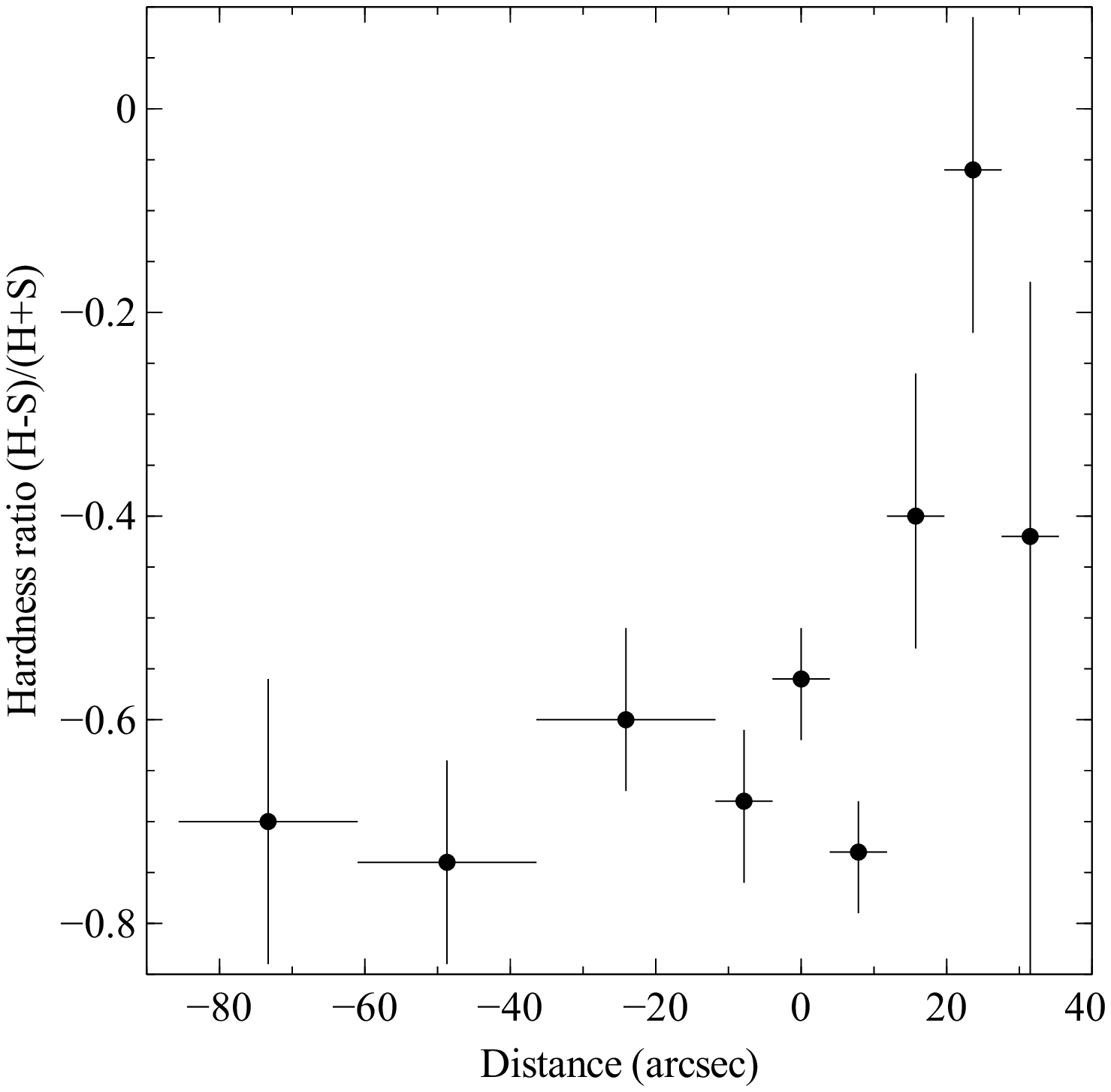}
\caption{Hardness ratio profile using a soft energy band $0.5-1.5\keV$
  ($S$) and a hard energy band $1.5-5.0\keV$ ($H$).  The profile is centred on the E elliptical galaxy in the N group and the W galaxy is at located $+22\asec$.}
\label{fig:hratio}
\end{figure}

Fig. \ref{fig:hratio} shows that the hardness ratio is roughly
constant at $-0.65$ through the NE group cool core then appears to
increase at the location of the `nose' where the W galaxy is located.
However, there are large uncertainties on the individual values.  This
could indicate a few keV temperature increase in the `nose' caused by
heating from the surrounding postshock gas.  Alternatively, there
could be a hard X-ray point source at the W galaxy centre consistent
with an AGN.  Although the FIRST survey does not show a radio point
source coincident with the W galaxy (\citealt{Becker95}), there also
does not appear to be radio emission associated with the hard X-ray
point source at the centre of the SW group.  The increase in hardness
ratio appears to extend over several neighbouring regions, which
suggests a temperature increase rather than a compact point source.

\section{Bow shock detection}
\label{sec:shock}

\begin{figure}
\centering
\includegraphics[width=0.98\columnwidth]{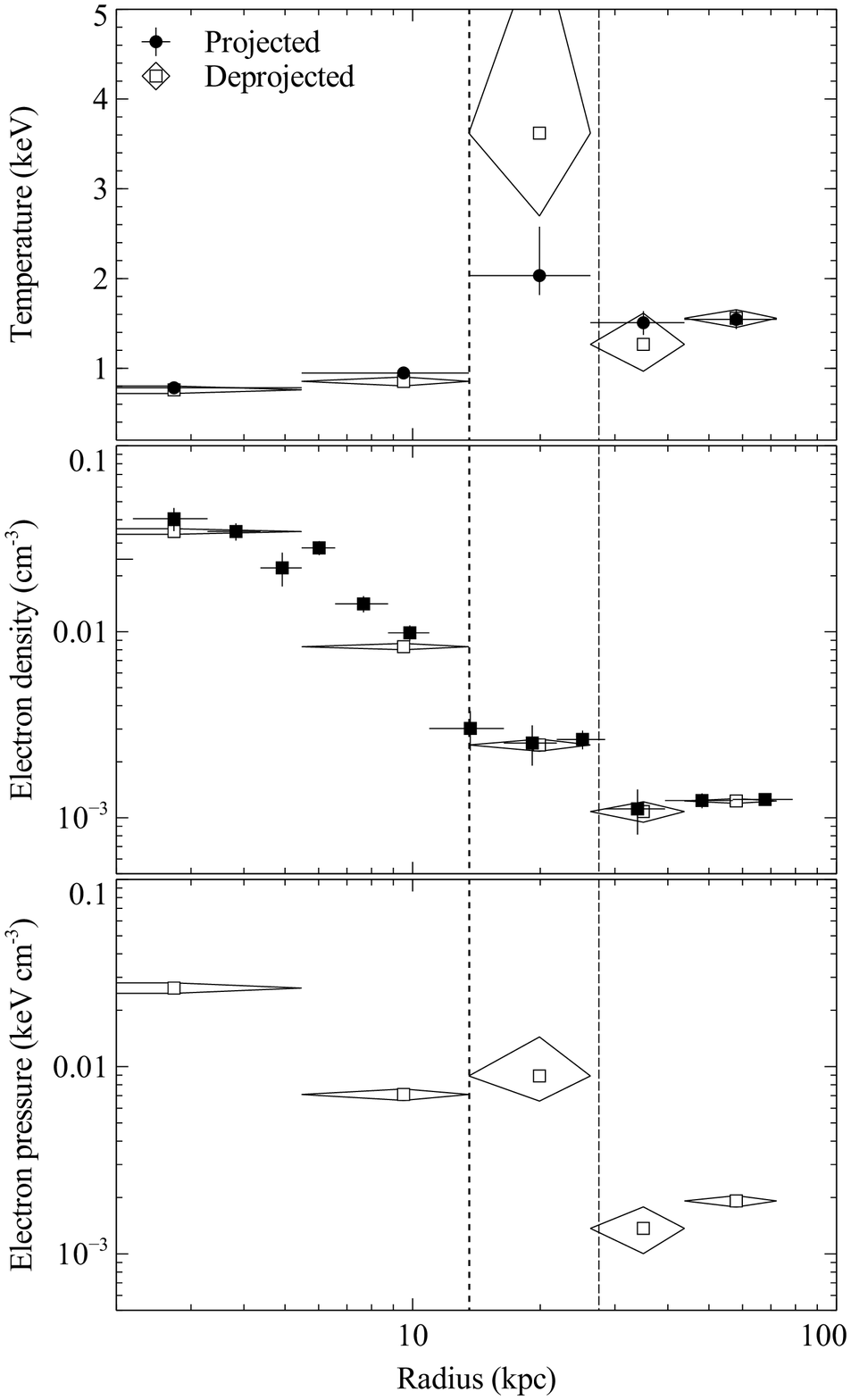}
\caption{Temperature (upper), electron density (middle) and electron
  pressure (lower) profiles for the N sector (see
  Fig. \ref{fig:SBsectors}).  The position of the surface brightness
  edge ahead of the N group core is shown by the dashed line.  Square
  symbols denote deprojected points.  The deprojected electron density
  profile was produced with \textsc{projct} (open squares) and a
  surface brightness deprojection (filled squares).  The dotted line
  at $14\kpc$ radius marks the merger cold front across the leading
  edge and the dashed line at $28\kpc$ radius marks the shock front.}
\label{fig:tempneshock}
\end{figure}



Fig. \ref{fig:SBsectors} (upper left) shows the N sector selected for
the analysis of the possible bow shock ahead of the NE group.  The
sector was positioned to align with the curvature of this outer
surface brightness edge and covers the angular range over which the
edge is clearly defined.  This sector was divided into 5 radial bins
that were positioned to determine the gas properties on either side of
the surface brightness edges with a minimum of 300 source counts per region.
Spectra extracted from these regions were each fitted with an absorbed
single temperature \textsc{apec} model, as detailed in Section
\ref{sec:spatialspec}.  The metallicity was fixed to the average of
$0.2\Zsun$ found for the N sector (Fig. \ref{fig:conbinmaps}).  The
projected temperature profile for the N sector is shown in
Fig. \ref{fig:tempneshock}.  The two central temperature bins at
$0.8-0.9\keV$ cover the NE group cool core.  The temperature increases
ahead of the core to $2.0^{+0.5}_{-0.2}\keV$ and then drops to
$1.5\pm0.1\keV$ beyond the outer surface brightness edge at a radius
of $27\kpc$.  The temperature and surface brightness, and therefore
the gas density, decrease significantly with radius across this outer
edge showing that it is a bow shock front ahead of the NE group core.


The \textsc{xspec} deprojection model \textsc{projct} was used to
subtract the projected contributions from the outer cluster layers off
the inner spectra and produce deprojected temperature and density
profiles.  Assuming spherical symmetry, \textsc{projct} simultaneously
fits to all the projected spectra accounting for projected emission by
adding suitably scaled model components from the outer annuli to the
inner spectral fits.  The X-ray emissivity depends strongly on the gas
density ($\propto{n^2}$) and only weakly on the temperature.  Therefore it is possible to produce
density profiles with higher spatial resolution by converting the
surface brightness to electron density and using modest corrections
for the temperature and exposure variation.
Fig. \ref{fig:tempneshock} shows a finer resolution deprojected
electron density profile overplotted on the \textsc{projct} result.
The electron density drops by a factor of $2.2\pm0.4$ across the bow
shock at a radius of $27\kpc$.  This coincides with a decrease in the
temperature by a factor of $3^{+2}_{-1}$.  Applying the Rankine-Hugoniot shock jump
conditions (eg. \citealt{LandauLifshitz59}), we can calculate the Mach
number of the shock, $M=v/c_s$, where $v$ is the preshock gas velocity
with respect to the shock surface and $c_s$ is the speed of sound in
the preshock gas.  The Mach number can be calculated from the density
jump,

\begin{equation}
M=\left(\frac{2\left(\ ^{n_{e,2}}/_{n_{e,1}}\right)}{\gamma + 1 - \left(\
    ^{n_{e,2}}/_{n_{e,1}}\right)(\gamma -1)}\right)^{1/2},
\label{eq:Machno}
\end{equation}

\noindent where $n_{e,1}$ and $n_{e,2}$ are the electron densities
upstream and downstream of the shock, respectively.  An adiabatic
index of $\gamma=5/3$, appropriate for a monatomic gas, is assumed.
The Mach number for the bow shock is therefore $M=1.9\pm0.4$.
Although the uncertainty on the postshock temperature is large, the
temperature jump appears broadly consistent with the density jump.  The sound speed in the preshock gas is $580\pm70\kmps$ and therefore the shock velocity is $1100\pm300\kmps$.

We used the shock velocity to estimate the time since the closest
passage between the two groups.  The galaxy redshifts suggest a line
of sight velocity difference of $\sim400\kmps$ between the two groups
(section \ref{sec:spatialspec}).  Therefore, we estimate that the
merger axis is $\sim10^{\circ}$ from the plane of the sky.  The
detection of a shock front is also consistent with a merger axis close
to the plane of the sky because projection effects would otherwise
obscure the surface brightness edge.  From the projected separation
between the two groups of $\sim90\kpc$, we estimate an age of
$\sim0.1\Gyr$.

The deprojected temperature and electron density were also multiplied
together to produce the electron pressure profile
(Fig. \ref{fig:tempneshock}).  This shows the expected sharp drop in
pressure across the bow shock front.  The pressure across the leading
edge of the NE group core is continuous within the error showing that
this is a merger cold front.  The pressure then peaks at the centre of the
NE group core.



\section{NE group core structure}
\label{sec:core}


\begin{figure}
\centering
\includegraphics[width=0.98\columnwidth]{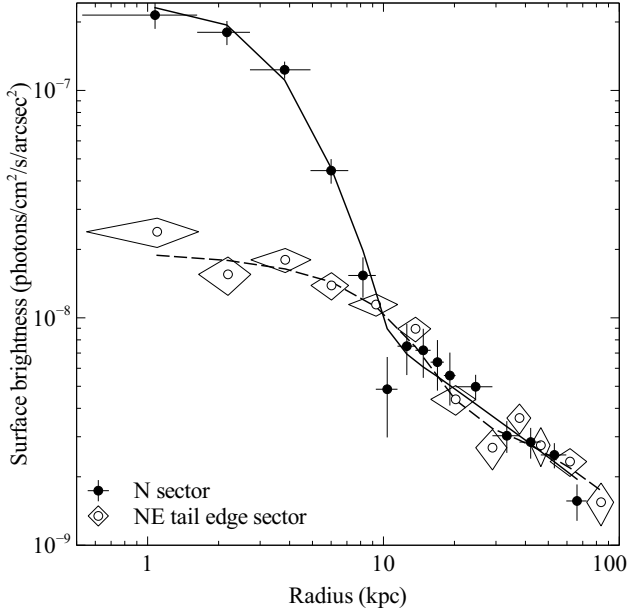}
\caption{Surface brightness profiles across the N sector (excluding the `nose') and the NE tail edge.  These regions are shown in Fig. \ref{fig:SBsectors} (upper left).  The best-fit projected density discontinuity models are shown overlaid (solid and dashed lines).}  
\label{fig:tailSBNedge}
\end{figure}

\begin{figure}
\centering
\includegraphics[width=0.98\columnwidth]{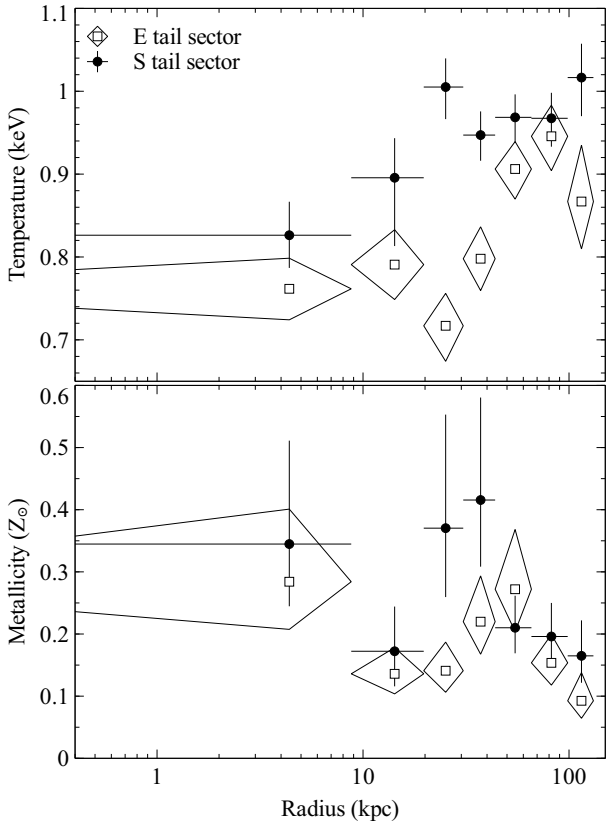}
\caption{Temperature (upper) and metallicity (lower) profiles for the NE group tail.  Note that these sectors are similar to the tail sectors shown in Fig. \ref{fig:SBsectors} but also extend over the centre of the tail.}
\label{fig:tailtempne}
\end{figure}

The N edge of the NE group core (Fig. \ref{fig:sbimage} left)
represents a sharp transition between the $1.5\keV$ ambient gas and
the cool core gas at only $0.7\keV$.  Fig. \ref{fig:conbinmaps} (upper
right) shows that this steep temperature gradient is preserved over
$\sim35\kpc$ distance behind the core along the NE edge of the ram
pressure stripped tail.  However, the tail appears to broaden rapidly
at a distance of $\sim45\kpc$ behind the dense core.  We
  compared a surface brightness profile across this edge with the
  surface brightness profile across the N edge
  (Fig. \ref{fig:SBsectors} upper right) to demonstrate this
  broadening (Fig. \ref{fig:tailSBNedge}).  The surface brightness
profile for the leading edge excluded the `nose'.  The profiles were
divided into radial bins of $5\asec$ width, increasing to $25\asec$ at
larger radii to ensure a minimum of 30 source counts per spatial bin.
The energy range was restricted to $0.3-4.0\keV$ to maximise the
signal-to-noise ratio.  The background was subtracted using the
normalized blank-sky background dataset (see Section
\ref{sec:datareduction}).  Fig. \ref{fig:tailSBNedge} shows the final
background-subtracted surface brightness profiles with the
characteristic shape of a projected density discontinuity.  The
leading edge of the dense core is particularly narrow and appears
broader in the sector along the NE edge of the subcluster's tail.

The surface brightness edges were fit with a model for a projected
spherical density discontinuity
(eg. \citealt{Markevitch00,Markevitch02}; \citealt{Owers09}).  The
radial density model consists of two power laws either side of a
density jump, which is then projected under the assumption of
spherical symmetry.  The model is then convolved with a Gaussian
function, to account for any intrinsic width, and fitted to the
surface brightness profiles.  The free parameters are the gradients
and normalizations of the power law models, the radius of the density
jump and the width of the density jump.  The best-fit model for the
leading edge and the NE tail edge are shown overlaid on the surface
brightness profiles in Fig. \ref{fig:tailSBNedge}.  The leading edge
of the NE group core has a best-fit width of $2.4\pm0.7\kpc$.  We note
that this is an upper limit as any variations in the shape of the edge
over the sector used will broaden the profile.  The edge widens to
$6\pm1\kpc$ in the sector across the NE edge of the subcluster's tail,
which coincides with the observed broadening of the temperature jump.
This is likely caused by the development of turbulent instabilities
along this sheared interface and will promote mixing of the cold ram
pressure stripped gas into the ambient medium.

\begin{figure*}
\begin{minipage}{\textwidth}
\centering
\includegraphics[width=0.48\columnwidth]{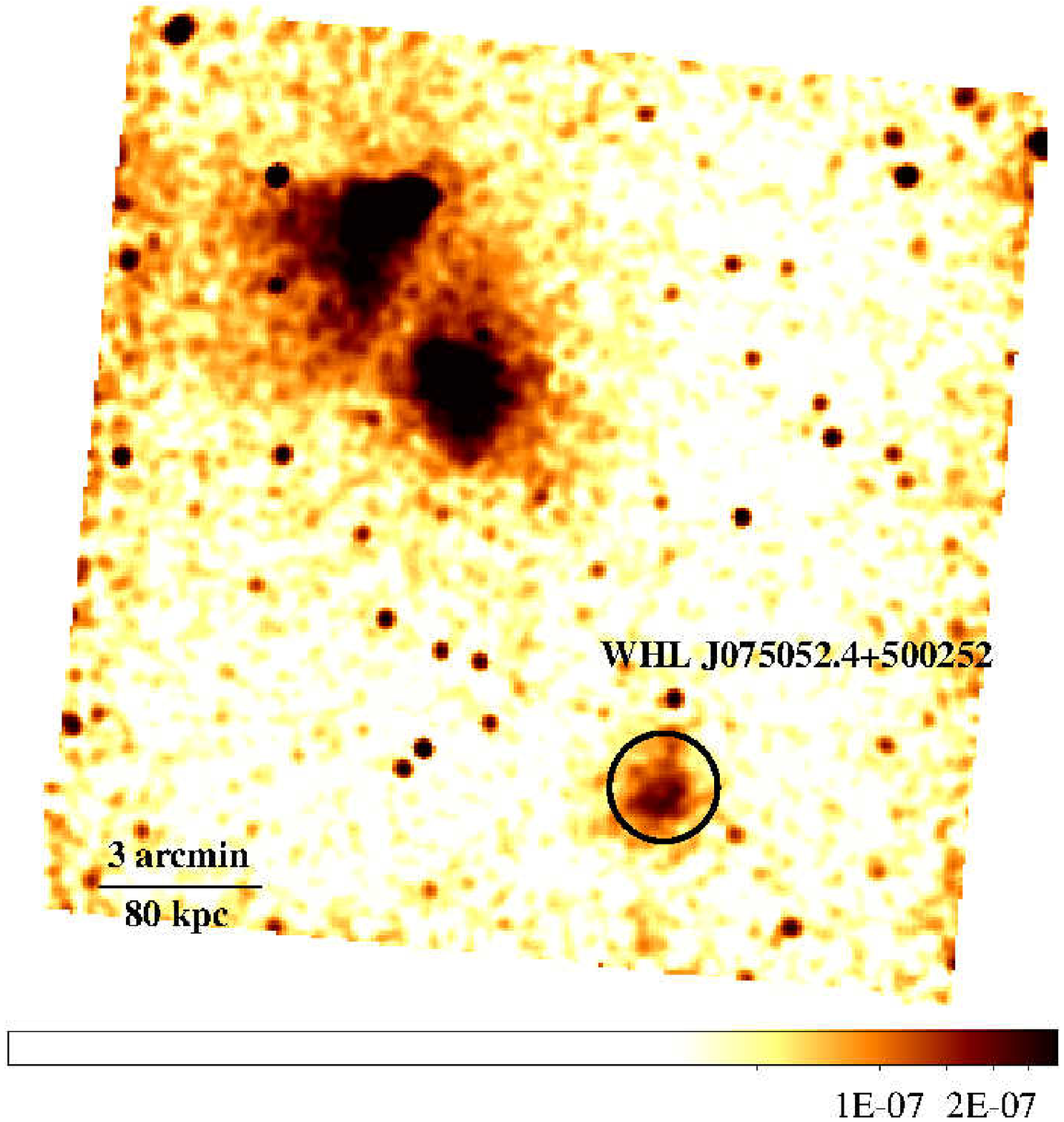}
\includegraphics[width=0.48\columnwidth]{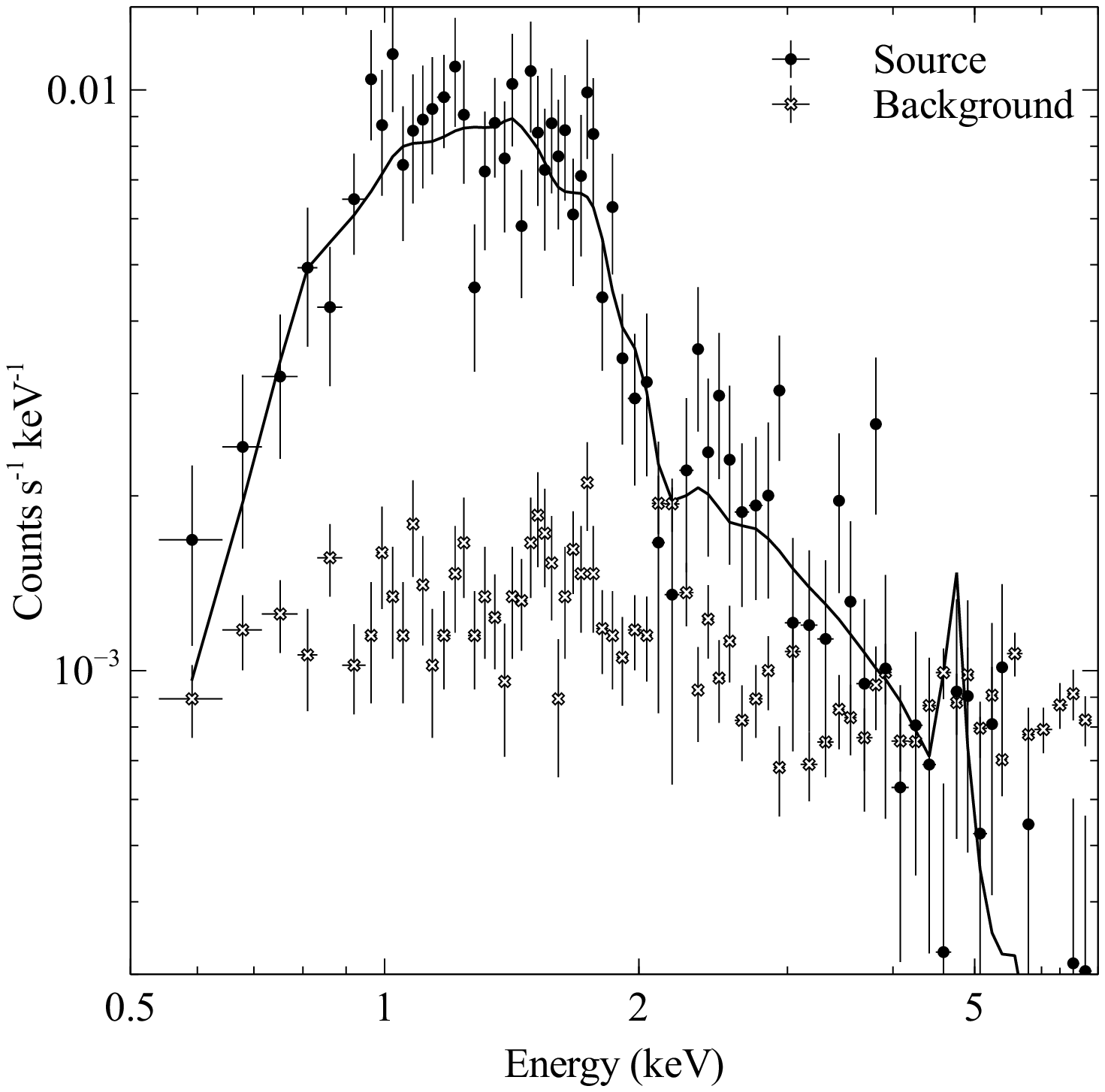}
\caption{Left: Exposure-corrected image showing the full field of the ACIS-I detector in the $0.3-4\keV$ energy band.  The extended emission from WHL J075052.4+500252 is visible on ACIS-I0.  The spectral extraction region is shown by the solid circle.  Right: Background subtracted spectrum for WHL J075052.4+500252 with best fit model overlaid (solid line).}
\label{fig:subcluster3}
\end{minipage}
\end{figure*}


The `nose' of the NE group core is likely to be a ram pressure stripped X-ray
corona hosted by the large elliptical galaxy UGC4052 W.  Following
\citet{Nulsen82}, the mass loss rate due to ram pressure is
$\dot{M}_{\mathrm{loss}}\sim{\pi}r^2\rho_{\mathrm{h}}v\sim0.3\Msunpyr$,
where $\rho_{\mathrm{h}}$ is the gas density in the ambient medium,
$r$ is the cross-sectional radius of the corona and $v$ is the gas
velocity.  Assuming a cylindrical volume and a density comparable to
the cool core, we estimate the total gas mass in the corona
$M\sim3\times10^7\Msun$ and therefore the survival time of the corona
is $\sim0.1\Gyr$ (see also \citealt{Sun05}; \citealt{Sun05Perseus}).
The corona may survive longer if there is significant mass injection
from the stellar population or ICM cooling (eg. \citealt{Acreman03})
or if the transport processes are suppressed by a magnetic field
(\citealt{Vikhlinin01Coma}; \citealt{Sun05}).


The striking IR and near-UV ring surrounding the galaxy pair UGC4052
is $\sim40\kpc$ across and the major axis appears aligned with the
central galaxies (Fig. \ref{fig:zoomin}).  This alignment suggests
that the formation of the ring could be linked to these galaxies.  The
group merger may have caused collisions between the constituent
galaxies and the ring could correspond to the orbit of a smaller
galaxy being stripped of its gas in the interaction.  However, there
is not a clear progenitor galaxy and the apparent spiralling of the
orbit to the W complicates this interpretation
(Fig. \ref{fig:zoomin}).  The ring may have been produced by the
stripping of multiple galaxies around UGC4052.  An explosion origin
appears less likely as the centre of the ring falls between the pair
of galaxies but cannot be ruled out.  Another possibility is that the
ring was produced by a close encounter between the two central
galaxies (eg. \citealt{Lynds76}; \citealt{Theys76};
\citealt{Appleton87,Appleton96}).  The passage of the companion galaxy
close to the centre of the parent galaxy produces an outwardly
travelling density wave in the gas and stellar disks and triggers star
formation in the expanding ring.  However, the two galaxies are
aligned along the major rather than the minor ring axis in contrast to
other collisional ring galaxies, such as the Cartwheel ring galaxy
(eg. \citealt{Higdon95}), Auriga's Wheel (\citealt{Conn11}) and
AM1724-622 (\citealt{Wallin94}).  There is a velocity difference of
$\sim250\kmps$ along the line of sight between the two galaxies
(\citealt{Crawford99}) and, if this interaction was triggered by the
group merger, there could be a greater velocity difference in the
plane of the sky.  UGC4052 E was targeted in the SDSS (\citealt{Abazajian09}) and the
spectrum shows a possible velocity offset between the gas emission lines
and absorption lines in the stellar continuum.  There could therefore also be a difference
in the gas and stellar dynamics in this complex system.

Fig. \ref{fig:zoomin} shows that at least part of the ring, depending on
the projection, is located in the shock-heated ambient medium.  PAHs,
which produce the strong emission feature at $7.7\mum$, are easily
destroyed by ionizing UV and X-rays and by shocks and therefore must
be shielded somehow in this environment (eg. \citealt{Voit92}).

\section{Ram pressure stripping of the NE group tail}
\label{sec:tail}
Fig. \ref{fig:tailimage} shows distortions around the sheared sides of
the NE group's ram pressure stripped tail and a potential large scale
hydrodynamic instability (eg. \citealt{Heinz03};
\citealt{Iapichino08}).  The shear flow around the NE group core will
be greatest along the S edge where the two groups are moving past each
other and the speed of the ambient medium is maximum
  (eg. \citealt{Mazzotta02}).  Fig. \ref{fig:conbinmaps} shows that
there is increased temperature and pressure in this region between the
two groups.  We compare the E and S sectors of the ram pressure
stripped tail to determine if there is potentially mixing between the
cool, dense core gas and the hot, ambient medium
(Fig. \ref{fig:tailtempne}).  The temperature profiles increase
through both sectors from $\sim0.75\keV$ to $\sim1.0\keV$, with a
higher temperature in the S sector out to $65\kpc$.  The temperature
in the S sector increases significantly at a radius of $25\kpc$ to
$1.01^{+0.03}_{-0.04}\keV$, which is comparable to the ambient gas.
This corresponds to the surface brightness depression identified in
Fig. \ref{fig:tailimage} (see Section \ref{sec:radprofiles}).  The
surface brightness excess in the neighbouring region at a radius of
$37\kpc$ has a temperature consistent with the NE group tail.
The excess is therefore a colder, denser structure extending
  into the ambient medium while the depression is a hotter, lower
  density filament penetrating from the ambient medium into the cool
  core.  The temperature and surface brightness structure are
  therefore consistent with the shearing along this edge producing
  turbulent mixing of the gas between the two merging groups.


Turbulent mixing is the most natural explanation for this structure
observed in a group merger and similar features have been found in
other merging systems.  The dense, cool cores in the cluster mergers
in A3667 and A2146 show comparable ram pressure stripped structures and
turbulent instabilities generated by the shear flow
(\citealt{Mazzotta02}; \citealt{Russell12}).  Observations of galaxies
moving through the ICM also show long tails of stripped gas, which can
be distorted by Kelvin-Helmholtz instabilities
(\citealt{Machacek05,Machacek06}; \citealt{Randall08M86};
\citealt{Sun10}; \citealt{Kraft11}).  The growth of these
instabilities can be suppressed by the surface tension of a magnetic
field and the viscosity of the gas, which may explain why some cold
fronts appear smooth and undisrupted (eg. \citealt{Machacek05}).
Gravity is not expected to be important for suppressing
Kelvin-Helmholtz instabilities in these systems
(eg. \citealt{Chandrasekhar61}; \citealt{Roediger12Groups}).  The
presence or absence of distortions to cold fronts can therefore be
used to constrain the local ICM viscosity and magnetic field strength
in the ICM.


Tangled magnetic fields may be amplified and stretched by the gas flow
around cool cores in a merger (eg. \citealt{VikhlininBfield01};
\citealt{Dursi08}).  Kelvin-Helmholz instabilities can be suppressed
by this magnetic layer that is aligned with the interface and the flow
direction.  Therefore, observations of instabilities put an upper
limit on the magnetic field strength at this location in
RXJ0751.3+5012.  Following \citet{Vikhlinin02}, the criterion for a
cold front \textit{stabilised} by the magnetic pressure is

\begin{equation}
\frac{B^{2}_{\mathrm{c}}+B^{2}_{\mathrm{h}}}{8\pi}>\frac{1}{2}\frac{\gamma{M^2}}{1+T_{\mathrm{c}}/T_{\mathrm{h}}}P_{\mathrm{ICM}},
\end{equation}

\noindent where $B_{\mathrm{c}}$ and $B_{\mathrm{h}}$ ($\muG$) are the
magnetic field strengths in the cold and hot gas either side of the
discontinuity, $T_{\mathrm{c}}$ and $T_{\mathrm{h}}$ are the
corresponding gas temperatures, $M$ is the Mach number for the gas
flow and $P_{\mathrm{ICM}}$ 
is the ambient pressure.  We
assume the adiabatic index $\gamma=5/3$ for a monatomic gas.
Therefore, for a maximum postshock gas velocity with $M\sim0.6$, gas
temperatures either side of the discontinuity of
$T_{\mathrm{h}}=2.0\keV$ and $T_{\mathrm{c}}=0.8\keV$, and
$P_{\mathrm{ICM}}\sim0.01\keVpcmcu$, the total magnetic field at this
discontinuity is less than $9\muG$.  This is consistent with
observational constraints from Faraday rotation measurements for other
galaxy groups (eg. \citealt{Laing08}; \citealt{Guidetti10}).



\citet{Roediger12} show that viscosity is not expected to
significantly suppress the development of Kelvin-Helmholtz
instabilities in galaxy groups.  Following \citet{Roediger13} and assuming Spitzer-like viscosity, we calculate the critical perturbation length scale below which viscosity will increase the growth time of perturbations

\begin{equation}
\begin{split}
\lambda_{\mathrm{crit}} =& 5\kpc\left(\frac{U}{500\kmps}\right)^{-1}\left(\frac{n_{\mathrm{e}}}{3\times10^{-3}\pcmcu}\right)^{-1} \\
 &\quad\times\left(\frac{k_{\mathrm{B}}T_{\mathrm{h}}}{2\keV}\right)^{5/2}
\end{split}
\end{equation}

\noindent where $U$ is the shear velocity and $n_{\mathrm{e}}$ is the
ambient gas density.  Therefore, for RXJ0751.3+5012, we estimate that
full Spitzer viscosity will suppress the growth of Kelvin-Helmholtz
instabilities on length scales below $5\kpc$.  This appears consistent
with the observed distortions on scales of $\sim10\kpc$ around the NE
group core but there may be smaller scale instabilities that cannot be
detected significantly in the existing observation.




\section{WHL J075052.4+500252}
\label{sec:highzcluster}

Fig. \ref{fig:subcluster3} (left) shows the full ACIS-I field of view.
The two merging subgroups of RXJ0751.3+5012 are visible to the NE on
ACIS-I3 but there is also extended emission clearly detected above the
background to the SW on ACIS-I0.  The X-ray emission peak for this
additional extended source is located at RA 07h50m52.4s Dec +50d02m55s
(J2000).  This corresponds to the galaxy cluster WHL J075052.4+500252
at $z=0.417$ (\citealt{Hao10}; \citealt{Wen12}).
Fig. \ref{fig:subcluster3} (right) shows the background subtracted X-ray spectrum
for this source and the background spectrum.  Using an
absorbed single temperature \textsc{apec} model, we found a best-fit
temperature of $5.2^{+1.3}_{-0.6}\keV$ and total X-ray flux of
$2.4^{+0.2}_{-0.1}\times10^{-13}\ergpcmsqps$ ($0.05-50\keV$).  The metallicity
was fixed to $0.3\Zsun$ and the redshift was fixed to $z=0.417$.
There was no clear detection of the Fe K line to confirm the redshift.
The best-fit values of temperature and luminosity ($L_X=1.49^{+0.12}_{-0.06}\times10^{44}\ergps$ for $0.05-50\keV$) place WHL
J075052.4+500252 on the $L_X-T$ relation within the errors
(\citealt{Markevitch98}; \citealt{Arnaud99}; \citealt{Pratt09}).


\section{Conclusions}


The new \textit{Chandra} observation of RXJ0751.3+5012 has revealed an
off-axis merger between two galaxy groups.  The NE group hosts a
dense, cool core that has been strongly disrupted by ram pressure in
the collision and formed a $\sim100\kpc$-long tail of stripped gas.
Sharp surface brightness edges are observed around each group core
corresponding to cold fronts where the cooler, dense core gas moves
through the warmer ambient medium.  The boxy structure and lack of
disruption to the SW group suggests the cold front around the core is
due to gas sloshing recently triggered by the collision.  We observe
an additional surface brightness edge $\sim15\kpc$ ($30\arcsec$) ahead
of the NE group core associated with a sharp drop in gas density,
temperature and pressure.  This is a bow shock front, the first
detected in a group merger, with a Mach number $M=1.9\pm0.4$.  From
the shock velocity of $1100\pm300\kmps$ and constituent galaxy
redshifts, we estimate that the merger axis lies close to the plane of
the sky at an angle of $\sim10^{\circ}$.  Together with the projected
distance between the groups of $\sim90\kpc$, we determine that closest
approach occurred $\sim0.1\Gyr$ ago.  From the global temperature of
each group, we estimated that the SW group is three times more massive
than the NE group and therefore the collision is on the borderline
between a major and minor merger.

Although the leading edge of the NE group core is narrow, the sheared
sides of the core and stripped tail are broader and appear distorted
by Kelvin-Helmholtz instabilities.  The gas temperature through the
ram pressure stripped tail increases more rapidly along the S edge
where the two groups are moving past each other and the shear flow is
strongest.  In addition, we observe an arc-like surface brightness
excess and depression consistent with the development of a large scale
hydrodynamical instability along the S edge of the tail.  The growth
of these turbulent eddies can be suppressed by magnetic fields and
viscosity.  Therefore, we show that, for Spitzer-like viscosity, the presence of these
  instabilities is consistent with the critical perturbation length
  above which instabilities can grow and
  place an upper limit on the magnetic field strength in the intragroup medium.

The NE group hosts two large elliptical galaxies (UGC4052) separated
by a projected distance of $\sim15\kpc$ and surrounded by an IR and
near-UV ring with a diameter of $\sim40\kpc$.  The E galaxy is
coincident with the X-ray peak and luminous H$\alpha$ line emission
located at the centre of the group's cool core.  The W galaxy is
positioned ahead of the group core and hosts a ram pressure stripped
X-ray corona.  The surrounding ring could have been produced by
material stripped from a smaller galaxy orbiting around UGC4052 or it
may be a collisional ring generated by a close encounter between the
two large galaxies.  


\section*{Acknowledgements}

We thank the reviewer for helpful and constructive comments.  HRR and
BRM acknowledge generous financial support from the Canadian Space
Agency Space Science Enhancement Program.  HRR also acknowledges
support from a COFUND Junior Research Fellowship at the Durham
University Institute of Advanced Study.  HRR and ACF acknowledge
support from ERC Advanced Grant Feedback.  PEJN was supported by NASA
contract NAS8-03060.  BRM acknowledges generous support from the
Natural Sciences and Engineering Research Council of Canada.  ACE
acknowledges support from STFC grant ST/I001573/1.  MD acknowledges
partial support from NASA Chandra award GO1-12155B.  HRR thanks Trevor
Ponman, Graham Smith, Lindsay King and Rebecca Canning for helpful
discussions.

Funding for the SDSS and SDSS-II has been provided by the Alfred
P. Sloan Foundation, the Participating Institutions, the National
Science Foundation, the U.S. Department of Energy, the National
Aeronautics and Space Administration, the Japanese Monbukagakusho, the
Max Planck Society, and the Higher Education Funding Council for
England. The SDSS Web Site is http://www.sdss.org/.  The SDSS is
managed by the Astrophysical Research Consortium for the Participating
Institutions. The Participating Institutions are the American Museum
of Natural History, Astrophysical Institute Potsdam, University of
Basel, University of Cambridge, Case Western Reserve University,
University of Chicago, Drexel University, Fermilab, the Institute for
Advanced Study, the Japan Participation Group, Johns Hopkins
University, the Joint Institute for Nuclear Astrophysics, the Kavli
Institute for Particle Astrophysics and Cosmology, the Korean
Scientist Group, the Chinese Academy of Sciences (LAMOST), Los Alamos
National Laboratory, the Max-Planck-Institute for Astronomy (MPIA),
the Max-Planck-Institute for Astrophysics (MPA), New Mexico State
University, Ohio State University, University of Pittsburgh,
University of Portsmouth, Princeton University, the United States
Naval Observatory, and the University of Washington.

\bibliographystyle{mnras} 
\bibliography{refs.bib}

\clearpage

\end{document}